\documentclass[fleqn,usenatbib]{mnras}
\usepackage{newtxtext,newtxmath}
\usepackage[T1]{fontenc}

\DeclareRobustCommand{\VAN}[3]{#2}
\let\VANthebibliography\thebibliography
\def\thebibliography{\DeclareRobustCommand{\VAN}[3]{##3}\VANthebibliography}

\usepackage{graphicx}	
\usepackage{amsmath}	
\usepackage{siunitx}
\usepackage{multirow}
\usepackage{xspace}

\newcommand{\lya}{Ly$\alpha$}
\newcommand{\HI}{\ion{H}{i}}

\newcommand{\hmpc}{h^{-1}{\rm Mpc}}

\newcommand{\kms}{\;{\rm km}\,{\rm s}^{-1}}

\newcommand\cdunits{{\rm cm}^{-2}}

\newcommand{\gizmo}{{\sc Gizmo}}

\newcommand{\simba}{\mbox{\sc Simba}\xspace}
\newcommand{\pygad}{\mbox{\sc Pygad}\xspace}

\title[High-ionisation Oxygen Absorption in \simba]{High-ionisation oxygen absorption from the Warm-Hot Intergalactic Medium in  \simba}

\author[L. Bradley et al.]{
Lawrence Bradley,$^{1}$\thanks{E-mail: ljbradley@hotmail.com}
Romeel Dav\'{e},$^{1,2,3}$
Britton Smith$^{1}$,
Weiguang Cui$^{1}$,
and Daniele Sorini$^{1}$
\\
$^{1}$Institute for Astronomy, Royal Observatory, Univ. of Edinburgh, Edinburgh EH9 3HJ, UK\\
$^{2}$University of the Western Cape, Bellville, Cape Town 7535, South Africa\\
$^{3}$South African Astronomical Observatories, Observatory, Cape Town 7925, South Africa
}

\date{Accepted XXX. Received YYY; in original form ZZZ}

\pubyear{2021}

\begin{document}
\label{firstpage}
\pagerange{\pageref{firstpage}--\pageref{lastpage}}
\maketitle

\begin{abstract}
We examine the physical conditions, environments, and statistical properties of intergalactic \ion{O}{vi}, \ion{O}{vii} and \ion{O}{viii} absorbers in the \simba cosmological hydrodynamic simulation suite. The goal is to understand the nature of these high ionisation absorbers, and test \simba's surprising prediction that $\sim 70\%$ of cosmic baryons at $z=0$ are in the Warm-Hot Intergalactic Medium (WHIM) driven by jet feedback from active galactic nuclei (AGN).  By comparing a full-physics \simba run versus one with jets turned off, we find that jet feedback causes widespread heating that impacts the absorption morphology particularly of the higher ions.  However, the distribution of the physical properties of detectable absorbers are not dramatically affected. Higher ionisation absorbers probe hotter gas as expected, but in \simba all ions arise at similar overdensities (typically $\delta\sim20-30$), similar environments (predominantly filaments), and similar nearest-halo distances (typically $\sim2-3r_{200c}$).  \simba matches the observed \ion{O}{vi} column density distribution function (CDDF) fairly well, but under-predicts the CDDF preliminarily derived from two detected intergalactic \ion{O}{vii} absorbers. Predicted CDDFs are very similar at $z=1$ with or without jets, but show differences by $z=0$ particularly at the high-column end.  Despite some discrepancies, \simba reproduces available observations as well as or better than other comparable simulations, suggesting that \simba's widespread jet heating cannot be ruled out by these data.  These results offer hope that future X-ray and ultraviolet facilities could provide significant constraints on galactic feedback models from high-ionisation IGM metal absorbers.
\end{abstract}


\section{Introduction}\label{sec: Intro}

At low redshifts, a census of baryons reveals that only $\sim20-30$\% of baryonic mass lies within bound structures such as galaxies, groups, and clusters~\citep{1998ApJ...503..518F}.  A significant fraction resides in the intergalactic medium (IGM), but observations of \ion{H}{i} absorption in the Lyman-$\alpha$ forest indicate only $\sim25-30\%$ more baryons associated with such absorbers~\citep{2012ApJ...759...23S}.  The current census data indicates $39\pm12\%$ of the total baryonic mass density is still unaccounted for at low redshift \citep{2017AN....338..281N}. This is commonly known as the missing baryons problem.
        
Cosmological simulations have long suggested that the main reservoir of these missing baryons is the Warm-Hot Intergalactic Medium (WHIM) \citep{1999ApJ...514....1C,1999ApJ...511..521D,2001ApJ...552..473D}. The WHIM is a gaseous phase with a temperature in the range of $T\sim10^5-10^7$~K, and outside of galaxy halos with baryonic overdensities of $\rho_b/\overline{\rho}_b\la 100$, mostly in filaments~\citep[see for example][]{2010MNRAS.408.2051D, Cui2019,Tuominen2021}. After the epoch of reionisation, the intergalactic medium (IGM) had a temperature of $\sim 10^4$~K, owing to photoionisation from the metagalactic ultraviolet flux. There are two main mechanisms that elevate matter to WHIM temperatures: gravitational shocks, and galactic feedback processes \citep{2011ApJ...731....6S} which can owe to star formation or black hole growth.  During hierarchical structure formation, baryonic matter in the cosmic web falls into the growing potential wells of large-scale structure, the gravitational perturbations result in shock heating and the growth of WHIM gas \citep{2001ApJ...552..473D}.  Feedback processes are canonically believed to be relatively unimportant in the diffuse IGM \citep{2006ApJ...650..573C}, but the recognition that relativistic jets from AGN may significantly impact the thermal state of surrounding gas in order to enact quenching of star formation in massive galaxies~\citep[``radio mode" feedback][]{Croton_2006,Bower_2006,SomervilleDave_2015} complicates the picture, because such extreme velocities could potentially propagate energetic material well into the IGM~\citep{2020MNRAS.491.6102B}, thus contributing to the generation of the WHIM.

\citet{Gurvich_2017} and \citet{2020MNRAS.499.2617C} explored the impact of AGN feedback on the low-redshift IGM using the Illustris~\citep{Vogelsberger_2014,Genel_2014} and \simba~\citep{2019MNRAS.486.2827D} simulations, respectively.  Both found dramatic effects on the Ly$\alpha$ forest owing to AGN feedback, strongly lowering Ly$\alpha$ absorption relative to models where AGN feedback is neglected.  This helped to resolve the so-called Photon Underproduction Crisis~\citep[PUC;][]{Kollmeier_2014} in which models (without AGN feedback) over-predicted by $\sim\times 5$ the amount of Ly$\alpha$ absorption at low-$z$, employing (at the time) state-of-the-art estimates of the photoionising background~\citep{2012ApJ...746..125H}.  While more recent estimates of the background can account for $\sim\times 2$~\citep{Khaire_2019,Faucher_2020} of the discrepancy, \citet{2020MNRAS.499.2617C} found that solving the PUC in \simba required including AGN jet feedback, which had the surprisingly major effect of reducing the cosmic baryon fraction in Ly$\alpha$ absorbing gas from $\sim 40\%$ to $\sim 15\%$.  Commensurately, this feedback mechanism increased the WHIM baryon fraction from $30\%$ to $70\%$ at $z=0$. This investigation was extended to higher redshift by \cite{Sorini_2021}, who showed that AGN jets in \textsc{Simba} decisively increase the baryon mass fraction in the WHIM phase after $z\approx 2$, while leaving the thermal state of the IGM relatively unaffected at earlier times. This behaviour explained why, contrary to what happens at low redshift, previous numerical studies found that the \lya\ absorption statistics in the IGM is rather insensitive to AGN feedback prescriptions at $z>2$ \citep{Sorini_2018, Sorini_2020}.

One might then hope that the dramatic increase in the WHIM at low redshift should be observationally testable.  Unfortunately, detecting WHIM gas is a challenge.  Neutral fractions become very low at WHIM temperatures, so that Ly$\alpha$ absorption only traces such gas when large concentrations of H lie along the line of sight and generate so-called broad Ly$\alpha$ absorbers~\citep{Tepper-Garcia_2012}, but their contribution is highly uncertain.  The most promising approach so far is to look for \ion{O}{vi} absorption, as this is a strong line whose collisional ionisation peak is at $3\times 10^5$~K~\citep{Cen_2001}, and \citet{2012ApJ...759...23S} estimated that \ion{O}{vi} absorbing gas could additionally account for $\sim 15-20\%$ of cosmic baryons, with \citet{Danforth_2016} detecting 280 \ion{O}{vi} systems at $z\la 1$ as part of the Cosmic Origins Spectrograph Guaranteed Time Observing (COS-GTO) program.  However, \ion{O}{vi}'s ionisation peak is narrow in temperature, while \ion{O}{vii} and \ion{O}{viii} peak at $T\sim10^{5.5}-\SI{e6.5}{\kelvin}$ and $T\sim10^{6.1}-\SI{e6.8}{\kelvin}$, respectively, making them in principle better probes of the bulk of WHIM gas.  Annoyingly, their strongest transitions lie in the soft X-ray regime which requires deep observations using {\it Chandra} \citep{2002ApJ...573..157N} and {\it XMM-Newton} \citep{2007ApJ...669..990B}.  Still, a handful of IGM \ion{O}{vii} absorbers have been detected (see \citealt{2018Natur.558..406N} and \citealt{Ahoranta2021} for nearby detections), allowing a preliminary estimate of the associated WHIM component which could, within substantial uncertainties, close the low-$z$ baryon census~\citep{2017AN....338..281N}.

Besides being a major baryon reservoir, WHIM absorbers could also provide key constraints on AGN feedback, if they substantially change the WHIM content.  Hence high-ionisation oxygen absorption has been studied in an assortment of cosmological simulations. For \ion{O}{vi}, neither Illustris \citep{2017MNRAS.465.2966S} nor EAGLE \citep{2016MNRAS.460.2157O,Ahoranta2021} was able to fully reproduce the observed column density distribution function (CDDF) of \ion{O}{vi}, with both predicting too little absorption at high column densities \citep{2015MNRAS.448..895S,2016MNRAS.459..310R}. 
In contrast, IllustrisTNG successfully reproduced the \ion{O}{vi} CDDF \citep{2018MNRAS.477..450N}, and made testable predictions for \ion{O}{vii} and \ion{O}{viii}.  The primary difference in TNG is the introduction of a ``jet mode" kinetic AGN feedback at low black hole accretion rates.  \cite{2019MNRAS.488.2947W} compared EAGLE to the \ion{O}{vii} equivalent width (EW) distribution inferred from just two extra-galactic absorbers, showing encouraging agreement albeit with the large uncertainties that preclude strong constraints.  Hence it appears that AGN feedback could noticeably impact high ionisation IGM oxygen absorbers, and that current models are not entirely consistent with observations.

In this paper, we explore WHIM absorbers in the \simba simulations \citep{2019MNRAS.486.2827D}.  We are particularly interested in the role that \simba's AGN jet feedback, which \citet{2020MNRAS.499.2617C} and \cite{Sorini_2021} showed strongly impacts the WHIM baryon mass fraction, has on the statistics of WHIM tracers \ion{O}{vi} ($1032,\SI{1038}{\angstrom}$), \ion{O}{vii} ($\SI{21.60}{\angstrom}$) and \ion{O}{viii} ($18.967, \SI{18.973}{\angstrom}$).  To quantify this, we will use the \simba suite's AGN feedback variant runs, which are identical except for the inclusion of the various forms of AGN feedback.  We examine the density, temperature and metallicity traced by these absorbers, and show how these are altered by jet feedback.  We also check the large-scale environments of WHIM absorbers by separating them into knots, filaments, sheets and voids through the \textsc{Pweb} classification \citep{Cui2018}, showing that all absorbers (even \ion{O}{viii}) predominantly lie in filaments and sheets.  We will make predictions for CDDFs in each ion, and show that including jet feedback results in good agreement with the observed \ion{O}{vi} CDDF particularly at high columns, but worse agreement with the two confirmed IGM \ion{O}{vii} absorbers.   Our results elucidate and quantify how these oxygen lines trace the missing baryons within the \simba simulation, and set the stage for future explorations with upcoming observational facilities.

This paper is organised as follows:  \S\ref{sec: SIMBA} will discuss the \simba simulation suite used here. \S\ref{sec: Physical conditions} will present the physical conditions in which the WHIM tracers were detected. \S\ref{sec: evolution} will study the evolution of the WHIM tracers. \S\ref{sec: obvs properites} will review the observational properties of the WHIM tracers, and make comparisons to observations. Finally, we summarise our results in \S\ref{sec: conculusion}.    

\section{The \simba Simulations}\label{sec: SIMBA}

    \subsection{Input physics}
    
\simba is suite of cosmological hydrodynamic simulations run with the \texttt{GIZMO} \citep{2015MNRAS.450...53H} code using its mass-conserving Meshless Finite Mass solver. It models a $\Lambda$CDM cosmology concordant with the \citet{2016A&A...594A..13P}: $\Omega_m=0.3, \Omega_\Lambda=0.7, \Omega_b=0.048, H_0=\SI{68}{\kilo\metre\per\second}\textup{Mpc}^{-1}, \sigma_8=0.82, n_s=0.97$. 
The high accuracy when dealing with shocks under MFM is advantageous when exploring the effects of high-velocity jets on the IGM (see below). 
        
To model galaxy formation, \simba employs a series of sub-grids to model key physical processes. First, radiative cooling and photoionisation heating are included using the \texttt{GRACKLE-3} library \citep{2017MNRAS.466.2217S}. This encompasses the metal cooling and non-equilibrium evolution of the primordial elements. Thermal equilibrium is not assumed, but ionisation equilibrium is; for hydrogen the collisional ionisation rates match those in \citet{1997NewA....2..181A}, while the recombination rates are from \citet{1997MNRAS.292...27H}. \simba assumes a spatially uniform ionising background, specified by \citet{2012ApJ...746..125H}. This has been modified to account for the self-shielding prescription of \citet{2013MNRAS.430.2427R}. 
        
For $H_2$-based star formation rates (SFR) a modified version \citep{2016MNRAS.462.3265D} of the Krumholz and Gnedin (\cite{2011ApJ...729...36K}) sub-grid model, based on metallicity and local column density, is used. The SFR is then calculated using the Kennicutt-Schmidt Law, with $SFR=0.02\rho_{H_2}/t_{dyn}$ \citep{1998ApJ...498..541K}. To model chemical enrichment the evolution of 11 elements (H, He, C, N, O, Ne, Mg, Si, S, Ca, and Fe) is tracked from Type Ia supernovae (SNe), Type II SNe and Asymptotic Giant Branch (AGB) stars, with the yield tables from \citet{1999ApJS..125..439I}, \citet{2006NuPhA.777..424N} as described in \citet{2006MNRAS.373.1265O}, respectively. Galactic winds from star formation are modelled using decoupled two-phase winds, using scalings from mass outflow rate and wind velocity versus galaxy stellar mass based on the results from the Feedback in Realistic Environments zoom simulations \citep{Muratov_2015,2017MNRAS.470.4698A}. \simba also includes metal-loaded winds; wind particles withdraw metals from local particles to represent local enrichment from supernovae. 
        
    \subsection{Black hole growth}
    
The black hole accretion model developed for \simba uses a unique two-mode system, which is dependent on the temperature of the gas surrounding the black hole. For hot gas ($T>\SI{e5}{\kelvin}$) within the black hole kernel encompassing 256 neighbours, \simba employs the Bondi accretion model \citep{1952MNRAS.112..195B}. In this mode the accretion rate is given by 
        \begin{equation}
            \dot{M}_{Bondi} = \epsilon_m \frac{4\pi G^2M^2_{BH}\rho}{(\nu^2+c_s^2)^{3/2}}
        \end{equation}
where $\rho$ is the density of the hot gas, $\nu$ is the average velocity of the gas relative to the black hole, and $c_s$ is the speed of sound in the hot gas. This mode is appropriate for hot gas since it models gas inflow from a dispersion-supported spherical cloud. 
        
However, Bondi accretion does not account for angular momentum losses limiting the accretion. Thus, for cold gas \simba employs a model based on \citet{2017MNRAS.464.2840A}, which is referred to as a `torque limited accretion' model following the description in \citet{2011MNRAS.415.1027H}. Briefly, this sub-grid model is a more appropriate description for a cold disc that is supported by rotation, as it accounts for how instabilities within such discs drive mass inflow.  See \citet{2019MNRAS.486.2827D} for full details.  So, for black holes in \simba the total accretion rate is given by 
        \begin{equation}
            \dot{M}_{BH} = (1-\eta) \times (\dot{M}_{Torque} + \dot{M}_{Bondi}),
        \end{equation}
where $\dot{M}_{Torque},\dot{M}_{Bondi}$ are the inflow rates from the two modes, and the constant radiative efficiency of $\eta=0.1$ is assumed \citep{2002MNRAS.335..965Y}.

\subsection{Black hole feedback}
    
The black hole feedback model in \simba is central to this work. This is implemented in two subgrid models: kinetic feedback and X-ray feedback, with the kinetic feedback split into `radiative' and `jet' models.  We describe these in more detail, in order to provie a full understanding of the \simba runs used here.
        
\subsubsection{Kinetic feedback}
        
The two kinetic feedback modes broadly mimic the dichotomy seen in radio galaxies where at high Eddington ratios, AGN are seen to drive fast winds believed to originate owing to radiation pressure off the accretion disk, while at low Eddington ratios AGN drive relativistic jets \citep[see review by][]{2014ARA&A..52..589H}.  Since it cannot directly model accretion disk scales, \simba assumes scalings between black hole properties and the wind velocity.  In all cases, it is assumed that the outflows are mass-loaded such that they carry a momentum of $0.2\dot{L}/c$, where $\dot{L}=0.1\dot{M}_{BH}c^2$ is the radiative luminosity.  Finally, all kinetic outflows are ejected bipolarly, along the angular momentum vector of the BH kernel with gas particles therein randomly selected; thus the AGN kinetic outflow in practice corresponds (initially) to a cylindrical column of gas on a $\sim$kpc scale. The direction can vary over time as the inner region changes orientation, but typically is stable over at least tens to hundreds of Myr.

The first of the two kinetic modes is referred to as `radiative mode', which is active at high Eddington ratios ($f_{Edd}\equiv\dot{M}_{BH}/\dot{M}_{Edd}>0.2$). In this mode, winds are driven at velocities of $\sim{500-1500}\kms$, which have been calibrated using SDSS observations \citep{2017A&A...603A..99P}. Designated as wind particles, they consists of molecular and warm ionised gas. This gas is ejected cool, consistent with the temperatures of $\sim10^4$~k observed in H$\alpha$ outflows \citep{2017A&A...606A..96P}, with a velocity given by 
            \begin{equation}
                v_{w,EL} = 500 + 500(\log M_{BH} + 6)/3\;\;\kms.
            \end{equation}
            
As the Eddington ratio decreases to $f_{\rm Edd}<0.2$ the jet mode of the kinetic feedback starts to slowly engage. The outflow velocity increases while the Eddington ratio continues to decrease, as given by 
            \begin{equation}
                v_{w,jet} = v_{w,EL} + 7000\log(0.2/f_{\rm Edd})\SI{}{\kilo\metre\per\second}
            \end{equation}

This thus mildly increases the velocity towards lower $f_{\rm Edd}$ at first, and then quickly as $f_{\rm Edd}$ approaches $0.02$. The maximum velocity in jet mode is then $\sim\SI{8000}{\kilo\metre\per\second}$, for $f_{\rm Edd}\leq 0.02$. To match the observations that jets are synchrotron-emitting plasma \citep{2012ARA&A..50..455F}, the temperature of jet mode feedback is raised to the virial temperature of the halo which we take as $T_{vir}=9.52\times10^7(M_{halo}/10^{15}M_\odot)^{1/3}\SI{}{\kelvin}$ \citep{2005AdSpR..36..701V}.  

We will refer to `jet mode` AGN feedback as that which occurs at full jet power, i.e. when $f_{\rm Edd}\leq 0.02$.  Note that the assumption of constant momentum input means that the mass loading factor of the kinetic winds scales inversely with velocity, while the energy carried is proportional to it.  Thus the jet mode, despite carrying less mass, dominates the energy output of the AGN, and the maximum kinetic power released is $\sim 30\%$ of the total accretion power, which is consistent with observations~\citep{Whittam_2018}.

\subsubsection{X-ray feedback}
        
The X-ray feedback in \simba is designed to model the heating (and overpressurisation) of gas within the BH kernel.  It is motivated by and broadly follows the subgrid implementation in the zoom simulations of \citet{2012ApJ...754..125C}.  This feedback mode has a relatively modest impact on the galaxy population overall, but serves to fully quench galaxies by removing any residual cool gas~\citep{2019MNRAS.486.2827D}, and can have important effects in evacuating the inner regions of green valley galaxies to bring them into better agreement with star formation rate profiles~\citep{Appleby_2020} and galaxy colour bimodality~\citep{Cui2021}.

X-ray feedback is only activated if a black hole satisfies the conditions for full jet mode feedback, and scales with the gas fraction within the kernel $f_{\rm gas}\equiv M_{gas}/M_*$ as $\propto 0.2-f_{\rm gas}$; at $f_{\rm gas}>0.2$, the radiative losses are assumed to be too great and no X-ray feedback is applied. The gas surrounding the accretion disc is heated due to X-ray emission from the disc. For non-ISM gas, the temperature is directly increased in accordance with the heating-flux. For ISM gas (with hydrogen number density $n_H<0.13$~cm$^{-3}$, half the X-ray energy is directly applied, while the remainder is applied kinetically as a radially outward kick. This is to prevent nonphysical cooling in the low-resolution ISM.
    
    \subsection{\simba runs}
    
Since we are interested in the effects of AGN feedback on IGM absorption, we employ \simba's suite of `feedback variant' runs, in which individual AGN feedback modes are turned off one by one.  These are run in a cubic box of length $50h^{-1}\textup{\textup{Mpc}}$, with $512^3$ dark matter and an equal number of gas elements. 
The two feedback variants runs we consider are:
\begin{itemize}
    \item `s50' refers to full AGN feedback.
    \item `s50nojet' refers to a run with X-ray and jet AGN feedback removed.
\end{itemize}
We have checked that the \simba's fiducial $100h^{-1}\textup{Mpc}$ box gives similar IGM properties to the 's50' variant which has identical input physics, just as the galaxy properties are well converged~\citep{2019MNRAS.486.2827D}.  But for uniformity we will use only the $50h^{-1}\textup{\textup{Mpc}}$ variant boxes, which are all run from the same initial conditions and have all other input physics the same except for the feedback modes.  Besides s50 which has the full \simba physics, these variants do not reproduce observed galaxy demographics, since key feedback processes are absent.  They are instead intended as numerical experiments to isolate and quantify the impact of individual feedback modes, most notably AGN jet feedback, on IGM oxygen absorption.

The \simba suite also contains a model with only X-ray feedback off (with jets still on) called 's50nox', one where all AGN feedback is turned off ('s50noagn'), and one where all feedback is turned off including star formation winds ('s50nofb'). \citet{2020MNRAS.499.2617C} and \cite{Sorini_2021} showed that s50nox gives IGM properties very similar to s50, while s50noagn likewise is quite similar to s50nojet, demonstrating that X-ray and radiative AGN feedback have minimal effects on intergalactic gas; the dominant physical process in \simba that alters the IGM at $z\lesssim 2$ is AGN jet feedback.  In Appendix A we show this is also true for the absorber statistics of high-ionisation oxygen lines at $z=0$, indicating that the most enlightening comparison is between s50 and s50nojet. Hence for brevity we will omit the other models in the main text and focus only on the s50 and s50nojet runs.

\subsection{Generating mock Spectra}

We generate absorption line spectra for \ion{O}{vi}, \ion{O}{vii}, and \ion{O}{viii} by selecting random lines of sight (LOS) through the simulation volume and computing the optical depths along each LOS using \pygad\footnote{https://bitbucket.org/broett/pygad/src/master/} \citep{2020MNRAS.496..152R}.  \pygad is a simulation analysis toolkit that is natively particle-based, and includes a module to generate mock spectra of any desired ion, fit a continuum, convolve it with a line spread function, and fit Voigt profiles\footnote{We have compared spectra from \pygad to the {\sc yt-}based spectrum generation software {\sc Trident}. Qualitatively, the results are similar, but the different smoothing procedure between \pygad's native SPH smoothing and {\sc Trident's} AMR-based approach can yield different features in detail, particularly in denser regions.  We use \pygad\ here because it is significantly faster, is natively Lagrangian, and is more easily modifiable to suit our needs as we describe below.}.

To produce a mock spectrum of a given ion along a given line of sight (LOS), \pygad uses the following method. Based on an input ultraviolet background (UVB), a pre-computed \texttt{CLOUDY} \citep{2017RMxAA..53..385F} lookup table provides the ionisation fraction of each gas element whose smoothing length intersects with the LOS. This accounts for both photoionisation and collisional ionisation, and is interpolated to the snapshot redshift being used. Here we assume a spatially-uniform ionising background given by \citet{Faucher_2020}, which \citet{2020MNRAS.499.2617C} showed results in a good match to the \HI\ mean flux decrement at low redshifts in \simba.  The gas elements are shifted into velocity space, and the ion density is smoothed along the LOS using the cubic spline kernel matching what is used in \simba, into pixels of a desired velocity binning. Physical constants of the ion are then used to convert the column density to optical depth. Besides the optical depth, \pygad also returns the temperature, density, metallicity (in the given element), and peculiar velocity along the line of sight. Each of these quantities has been weighted by the optical depth, so they represent the physical conditions of the gas that is doing the absorption in that ion.

We generate spectra that broadly matches the characteristics of the COS-GTO survey~\citep{2016MNRAS.462.3265D}.  We choose $\approx 6\kms$ pixels, and add a signal-to-noise ratio per pixels of $S/N=20$, with Gaussian noise; this is somewhat higher than the typical $S/N$ in the COS-GTO data, but we would like to ensure completeness at low columns. Since the most main observational comparison will be to \ion{O}{vi} observations from COS-GTO, our procedure was primarily geared to mock this dataset.  Nonetheless, for uniformity, we also generate the \ion{O}{vii} and \ion{O}{viii} using the same spectral characteristics.  This is far better than what can be achieved with current X-ray telescopes, so represents a prediction for future facilities.

Each spectrum underwent a continuum fitting procedure as described in \citet{2020MNRAS.499.2617C}, although this typically results in negligibly small change from the true continuum given by the simulation.
For the \ion{O}{vi} spectra that were generated, a line spread function (LSF) for the COS G-130M grating was applied, as this is the grating primarily used in the COS-GTO dataset. These are the spectra that are then used for the analysis.

We then determine the physical properties of the absorption. First, we detect regions where there is significant absorption. To determine a detection region, the flux of a spectrum was smoothed using a variety of Gaussians, which have a standard deviation range of 2 to 11 pixels. After smoothing, for each pixel a detection ratio was determined. This is defined as the ratio between the convolved flux and the square root of the convolved noise. For each pixel, the standard deviation that yielded the highest detection ratio was selected as that pixel's detection ratio. The pixels were then looped over to find a contiguous region where the detection ratio for each pixel was above a given threshold, which we set to be 4; thus detection regions are required to have at least $4\sigma$ significance. The detection region is then expanded by several pixels to account for the wings.

Within each detection region, a Voigt profile (VP) fitter was applied.  The VP fitter broadly follows the procedure in {\sc AutoVP} \citep{Dave_1997}, but omitting the {\tt autofit} initial guess which was found to not particularly benefit the final fit.  \pygad's VP fitter (written by co-author Dav\'e) begins by fitting a single Voigt profile at the location of the minimum flux within each detection region, and then iteratively adds lines at the minimum residual flux so long as the deviation of the model fit from the input spectrum exceeds a reduced $\chi_\nu^2>2$ within that region.  The best-fit model is determined using {\tt scipy.optimize.minimize()}. The output of \pygad's VP fitting is a list of identified lines, with a wavelength, column density, and line width for each line.  An error estimate on these quantities is also output as the square root of the diagonal of the covariance matrix.

We associate physical conditions from the input spectrum to each absorption line.  To do this, we take the wavelength of the absorption line, find the closest pixel in the original spectrum, and use the physical conditions of that pixel.  The physical conditions are computed in \pygad\ along each LOS as the optical depth-weighted density, temperature, and metallicity for each ion (this part was written by co-author Sorini), so should reflect the conditions of the gas giving rise to that particular ion's absorption.  We note that this can be somewhat approximate, as the smoothing of the physical properties onto pixels along the LOS can introduce phase mixing.  Nonetheless it should be good enough to get a sense for the trends within the bulk of the absorber population.

We select 10,000 randomly-chosen lines of sight at each redshift along the $z$-axis of the simulation.  We do so at $z=0,1$ to study the differences in physical conditions among absorbers and their redshift evolution, as well as at intermediate redshifts as appropriate to compare to observations.

\subsection{Sample mock spectra}
    
        \begin{figure*} 
            \centering
            \includegraphics[width=\linewidth]{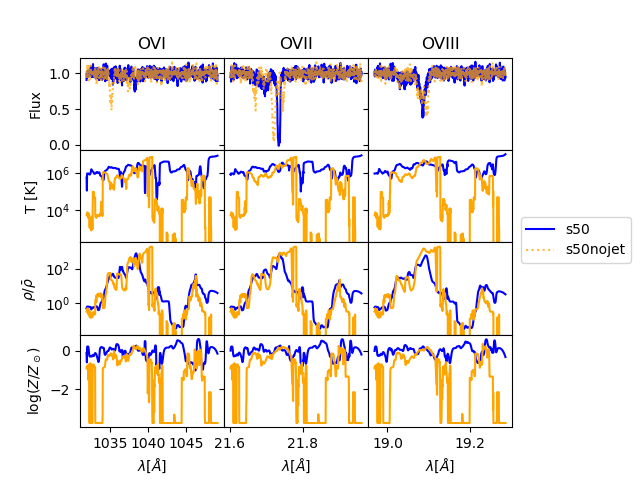}
            \vskip-0.1in
            \caption{An example $z=0$ line of sight generated using \texttt{PYGAD}, for \ion{O}{vi} (left column) \ion{O}{vii} (middle) and \ion{O}{viii} (right).  The same LOS is shown for the s50 (blue) and s50nojet (orange) runs. The flux (top), temperature (second row), overdensity (third row) and oxygen metallicity (bottom) are plotted in wavelength space. The physical quantities are weighted by the ion optical depth, so they differ slightly among the three ions.  Note that this LOS is not typical, as it shows some absorption in all three ions.}
            \label{fig: Example Spectra}
        \end{figure*}

Figure \ref{fig: Example Spectra} displays an example \pygad generated spectrum along a LOS for \ion{O}{vii}, in the s50 (blue) and s50nojet (orange) runs at $z=0$. The flux (top), the temperature (second row), overdensity (third row) and oxygen metallicity relative to solar (bottom) are all plotted versus wavelength.  We note that this is not a typical spectrum, but one that shows significant absorption in all ions.

Because the initial conditions of these runs are identical, the spectra probe the same overall large-scale structure through the volume.  Nonetheless, the spectra show notable differences in the absorption and physical quantities between s50 and s50nojet.  This illustrates that AGN feedback makes a significant difference to the predicted \ion{O}{vii} absorption, at least in this one case.  While the absorption broadly occurs in the same location along in the line of sight in the s50 and s50nojet cases, there are differences in the strength and exact positions of the lines.  This example has strong \ion{O}{vii} and \ion{O}{viii} absorption, but the corresponding \ion{O}{vi} absorption is weak, which highlights the typical situation that it is difficult to find regions that have enough diversity in density and temperature within a single absorbing structure to have strong absorption in all three ions.
        
The second panel shows strong temperature differences in some regions of the simulation box induced by AGN jets.  In the no-jet case, the temperature in the diffuse IGM is predominantly set by gravitational shock heating on large-scale structure.  This heats some regions to $\sim 10^6$~K.  But with jets on, virtually the entire line of sight is at these temperatures, indicative of widespread heating and the greater WHIM baryon fraction in the s50 case~\citep{2020ARA&A..58..363P,Sorini_2021}.  

The third panel shows the optical depth-weighted overdensity, which indicates a significant structure with overdensities exceeding $\ga 100$ that is indicative of gas within or near a halo.  The overdensity is mostly set by gravity that generates the cosmic web and thus is not as strongly impacted by jets, but it is clear that the peak overdensity can still be lowered due to the heating from AGN jets; this is consistent with the lowering of the baryon fraction out to many virial radii due to jets~\citep{Sorini_2021}.

The bottom panel shows the impact of jet feedback on metallicity in the simulation. In the s50 run there are more regions with a higher metallicity, and this is a clear sign that the jet feedback is responsible for creating more widespread metal enrichment in the IGM.  The metallicity value is around solar along the entire line of sight in the s50 case, which may seem high but it is likely arising due to the optical depth weighting which \citet{Oppenheimer_2009} points out significantly enhaces the metallicity relatively to the volume averaged value.

\subsection{Projected absorption maps}

        \begin{figure*}
            \centering
            \includegraphics[width=0.49\linewidth]{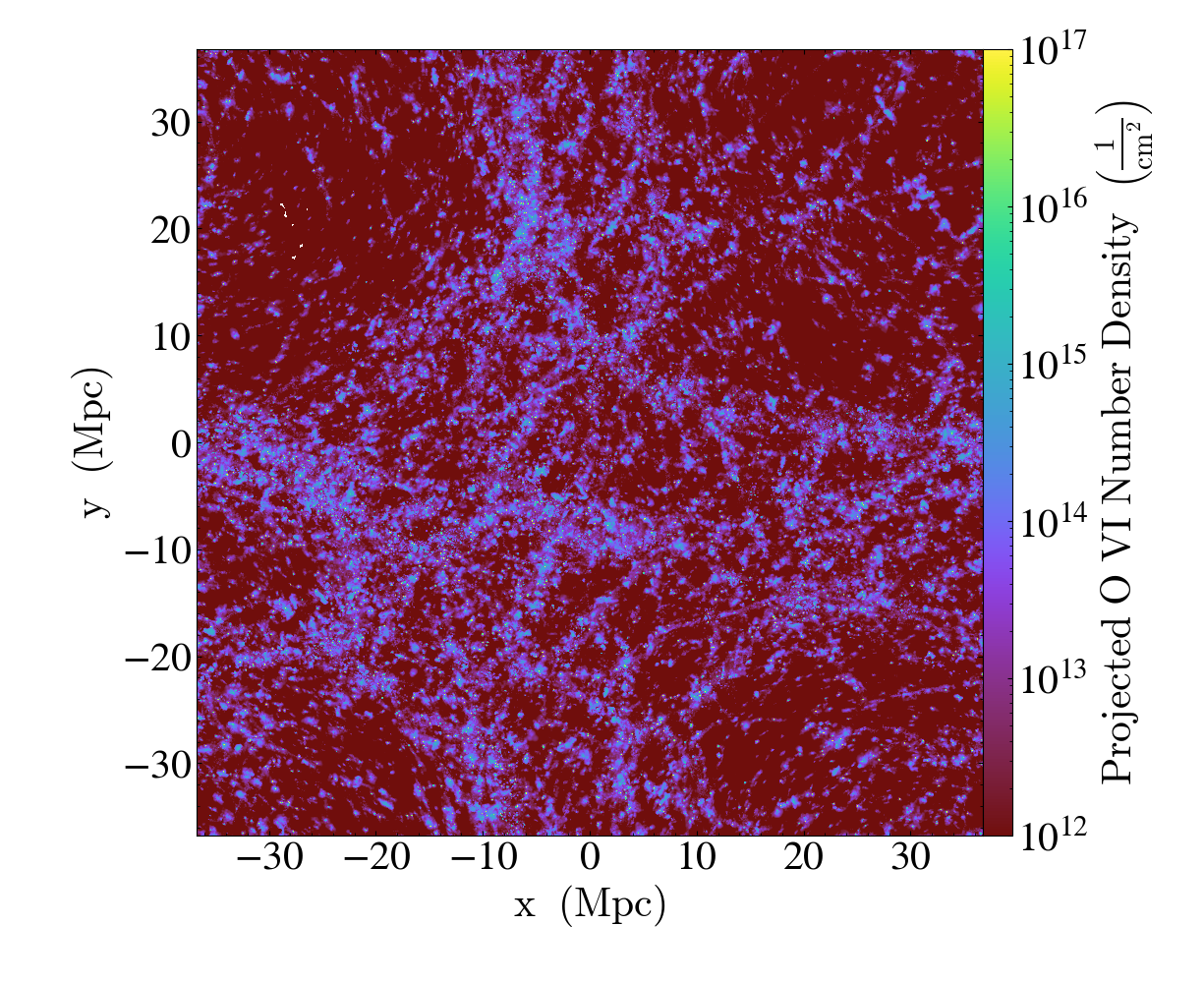}
            \includegraphics[width=0.49\linewidth]{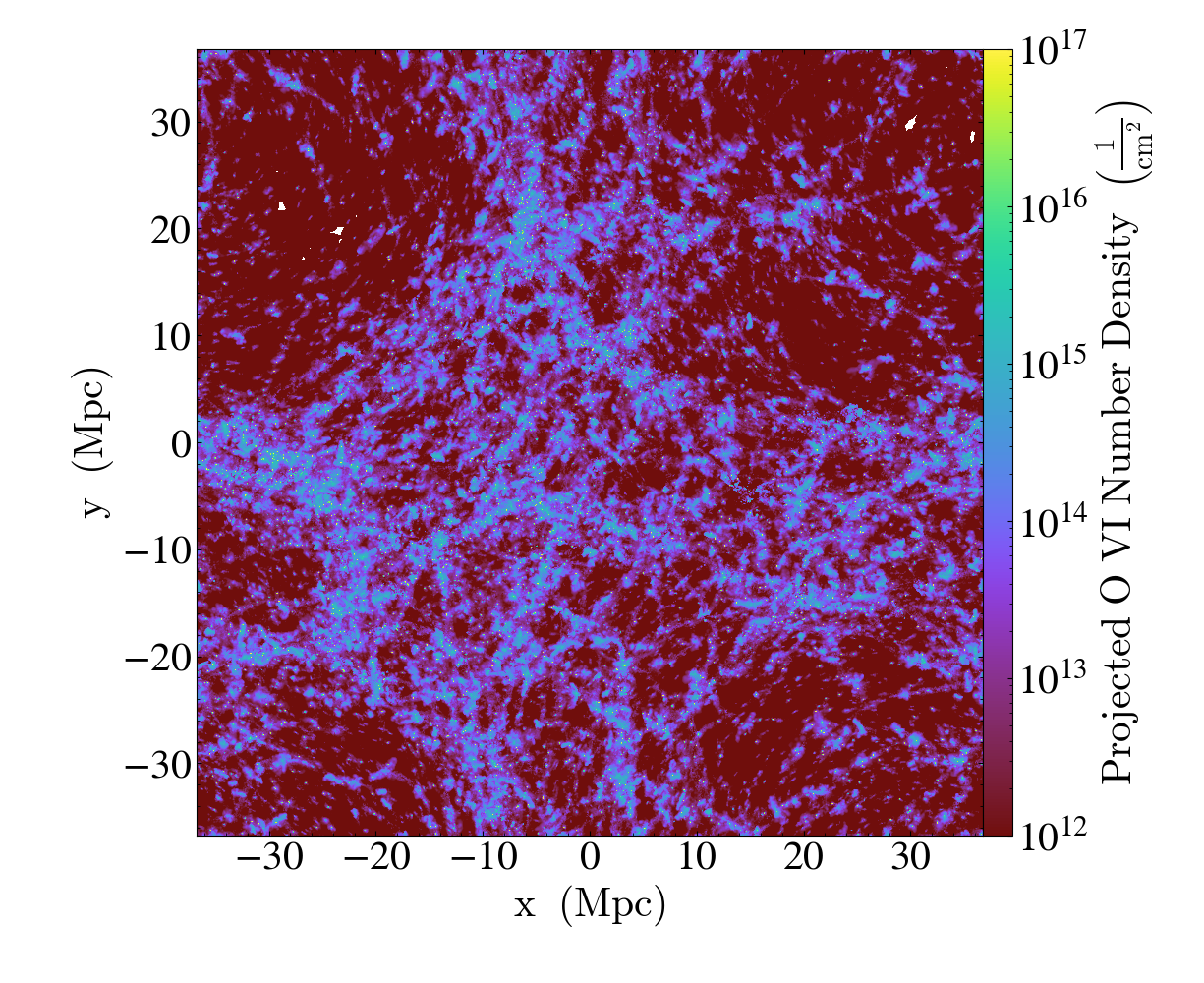}
            \includegraphics[width=0.49\linewidth]{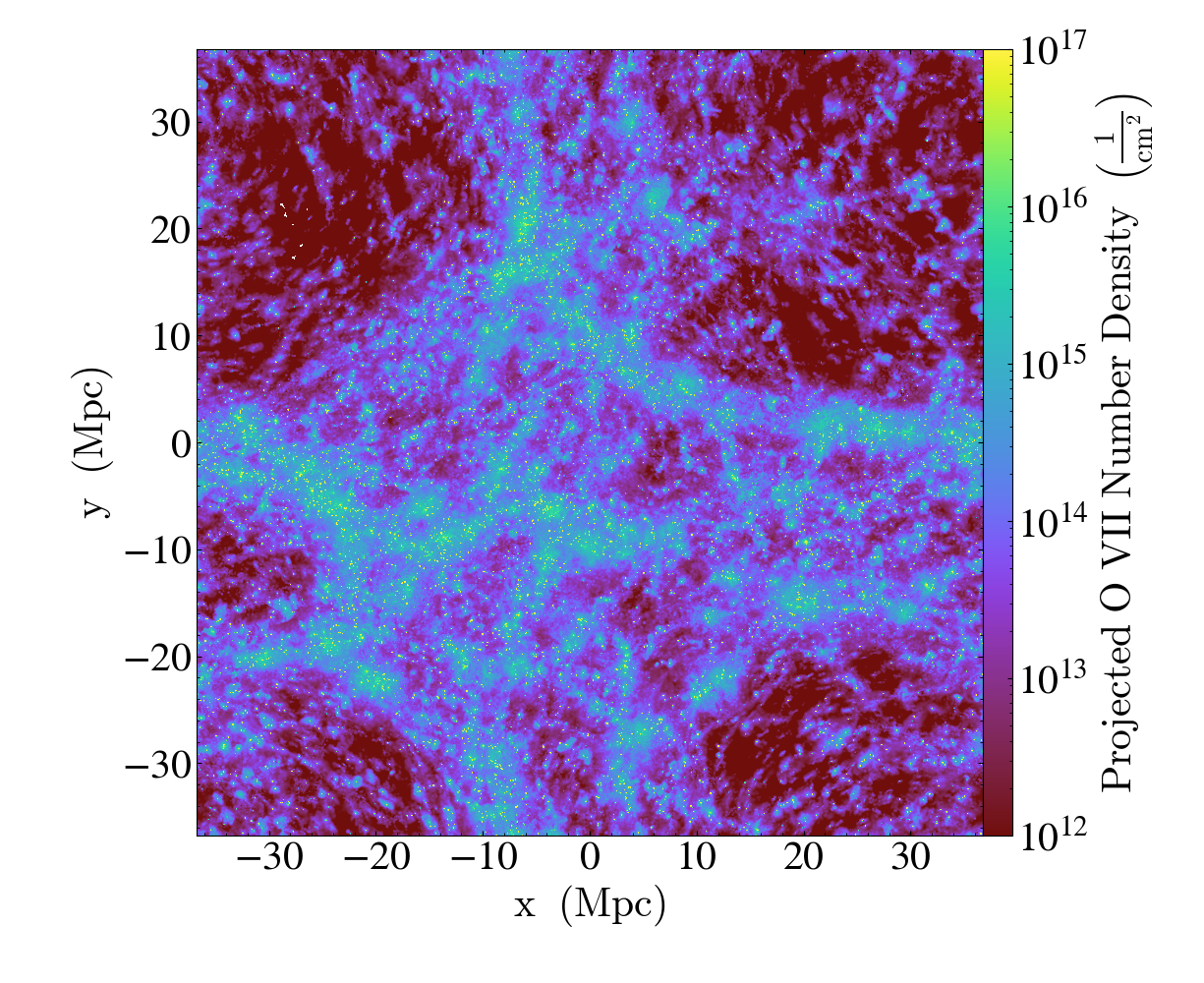}
            \includegraphics[width=0.49\linewidth]{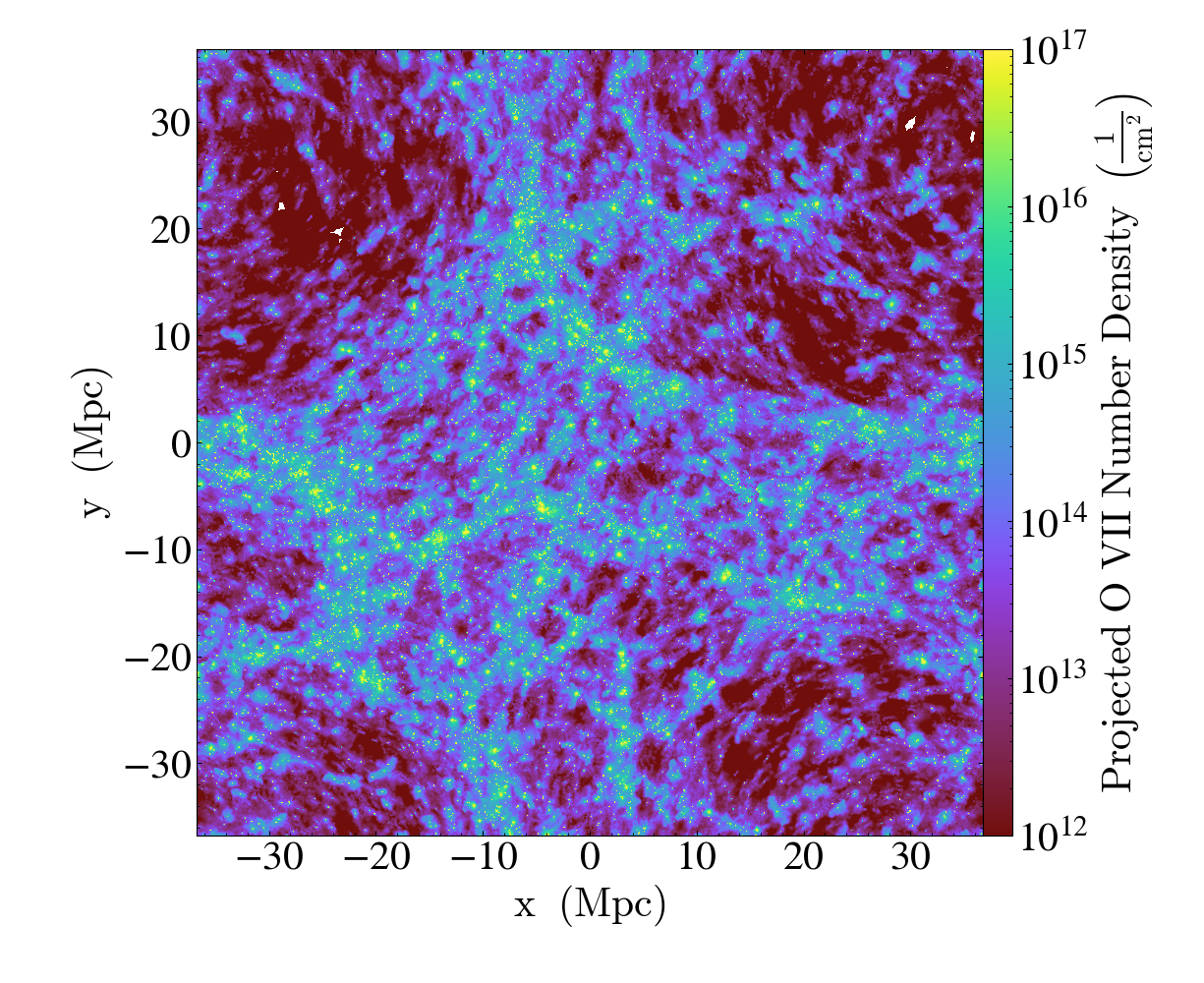}
            \includegraphics[width=0.49\linewidth]{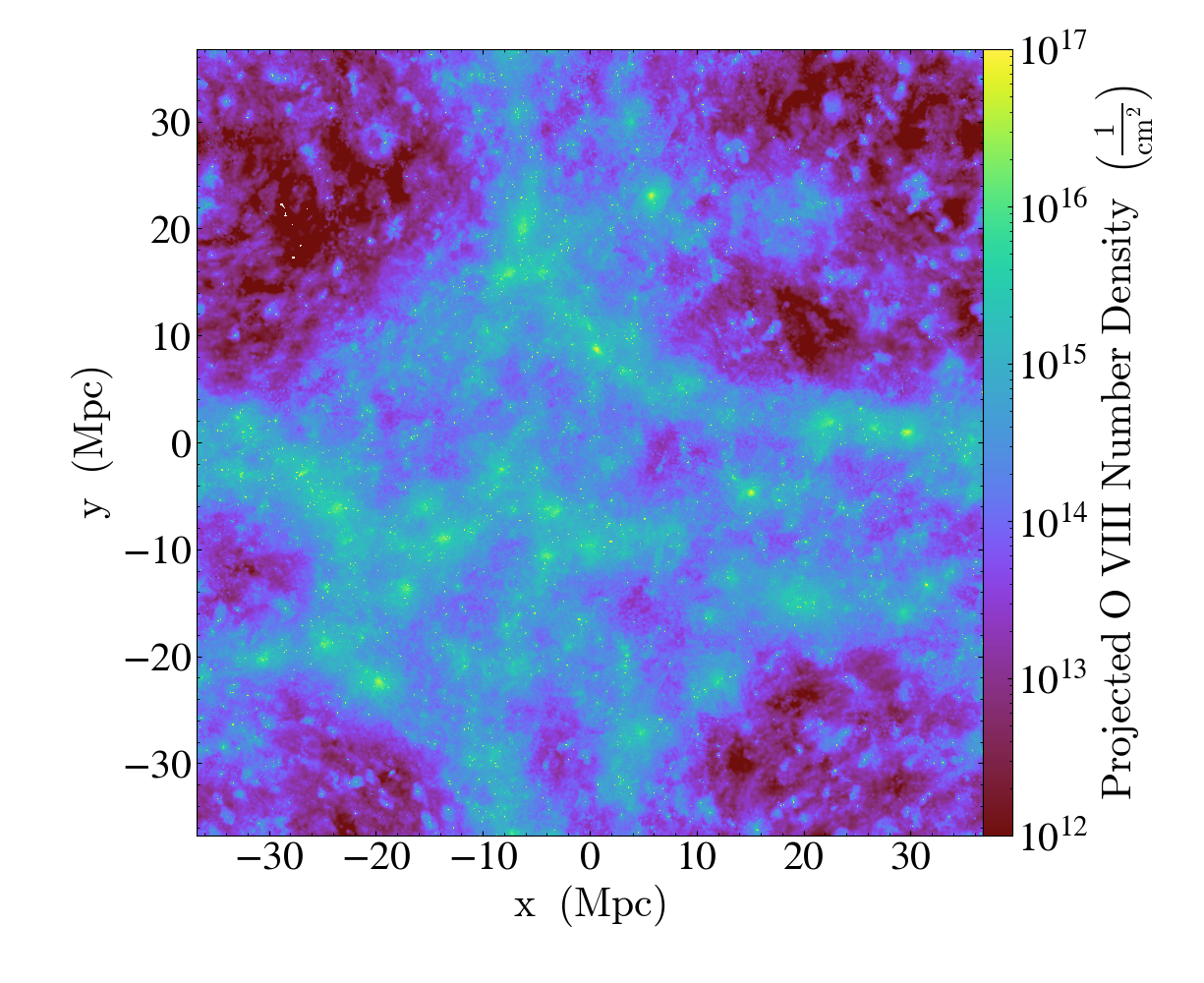}
            \includegraphics[width=0.49\linewidth]{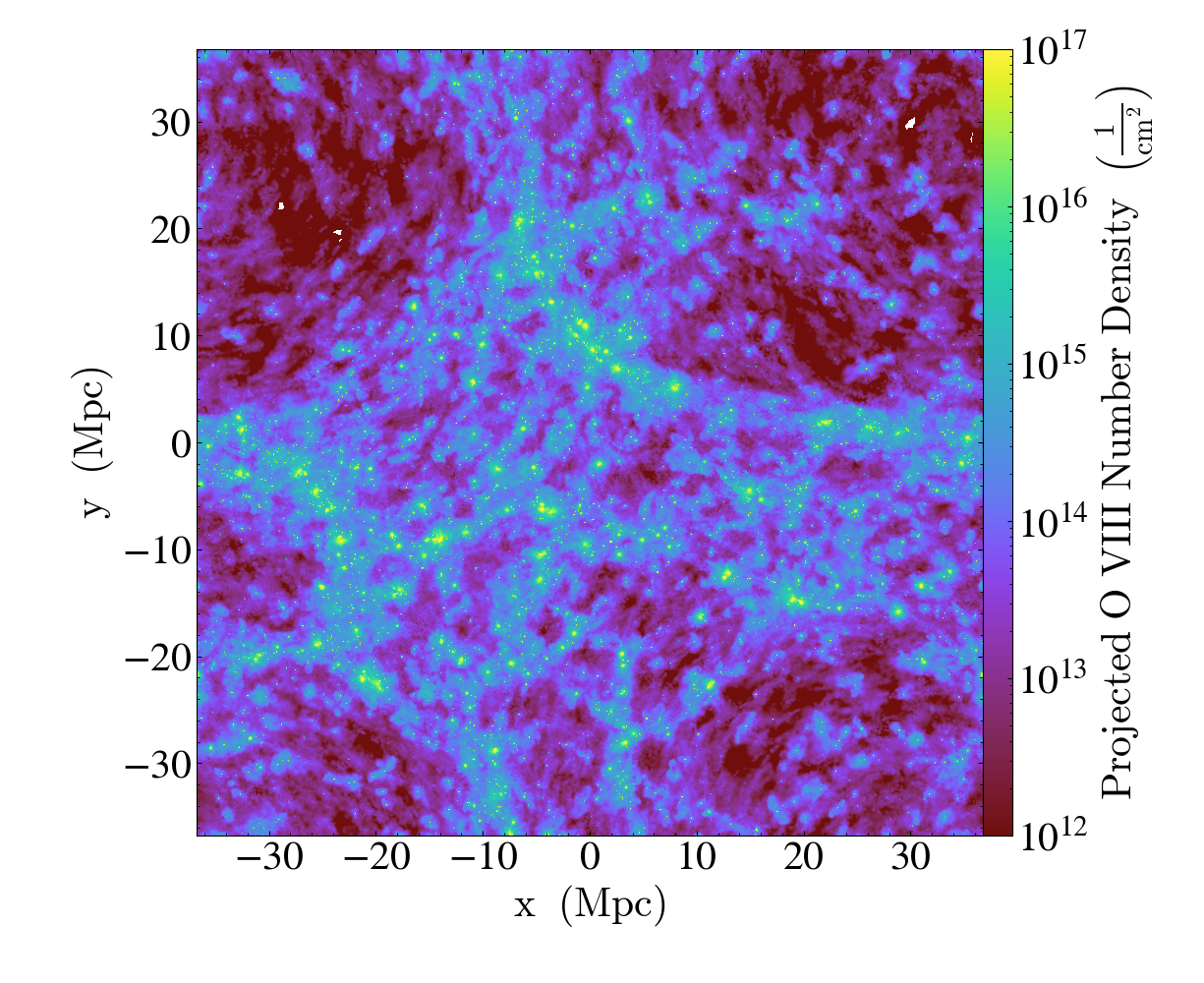}
            \vskip-0.1in
            \caption{Projected column density maps, for \ion{O}{vi} (top), \ion{O}{vii} (middle) and \ion{O}{viii} (bottom), generated at $z=0$ for the s50 (left column) and s50nojet (right column) runs.
            The full column density range is restricted.}
            \label{fig: maps}
        \end{figure*}
        
Figure~\ref{fig: maps} shows projected column density maps through the entire simulation volume at $z=0$ of our three ions, for the s50 run (left column) and s50nojet run (right column).  From top to bottom is shown \ion{O}{vi}, \ion{O}{vii}, and \ion{O}{viii}. The column density colour scale runs from $10^{12}$ to $10^{17}\cdunits$ in each case.  These maps were created using {\tt yt} \citep{2011ApJS..192....9T} and its absorption line extension package, {\tt Trident} \citep{2017ApJ...847...59H}.

All ions trace out the filamentary cosmic web present in these runs.  By $z=0$, these filaments contain substantial hot gas owing to shock heating on large-scale structure, as well as (in the s50 case) heating from AGN feedback.  \ion{O}{vi} is the weakest line, owing to the fairly narrow temperature range in which it is strong.  \ion{O}{viii} appears to trace the most gas overall into the diffuse regions, while \ion{O}{vii} is intermediate between these.  Note that this does not account for the current detectability of these various ions; \ion{O}{vi} is far more commonly observed at present owing to current far-UV spectrographs being able to trace absorbers with column densities below $10^{13}\cdunits$, while \ion{O}{vii} and \ion{O}{viii} lie in the soft X-ray regime where current X-ray telescopes require long integration times to identify lines below $10^{16}\cdunits$.

Comparing s50 and s50nojet, we see that for \ion{O}{vi}, s50 seems to have slightly less prominent absorption than s50nojet.  For the higher ions, the most notable feature is that the absorption is more ubiquitous in s50, reflecting the widespread heating of the jets.  \ion{O}{viii} in particular looks stronger in s50nojet versus s50.

We quantify these trends statistically by constructing column density distribution functions (CDDFs) in \S\ref{sec: obvs properites}.  For this, we will use these projected column density maps.  The reason is, even with 10,000 random lines of sight, the absorber population is relatively sparse, particularly at column densities that are feasibly detectable for \ion{O}{vii} and \ion{O}{viii}, so using the maps enables us to probe a wider dynamic range in column densities  We have checked versus the VP fitting results that using the full projection does not introduce a significant bias in the CDDF, because the incidence of multiple strong absorbers along a given LOS is fairly rare. The VP fitting results will be used primarily to examine the physical conditions of individual absorbers.

\section{Physical Properties of WHIM Tracers}\label{sec: Physical conditions}

We begin by examining the physical properties and environments of the \ion{O}{vi}, \ion{O}{vii}, and \ion{O}{viii} absorbers seen in our \simba runs.  We will discuss their location in phase space and their location in physical space, and how these are impacted by AGN jet feedback.  This will provide some context as to how to best interpret present and upcoming absorption line observations within the context of current galaxy evolution models.

\subsection{Location in Phase Space}

The abundance in a given ionisation state is governed by the temperature, density, and metallicity of the absorbing gas. In our simulation, we have direct access to these quantities, so we can study the physical conditions of different absorbing ions.  We first examine phase-space diagrams of our identified absorbers.  The density and temperature of each absorber is obtained from \pygad as the values at the pixels nearest to the fitted line's location in wavelength space.
    
    \begin{figure*}
        \centering
        \includegraphics[width=\linewidth]{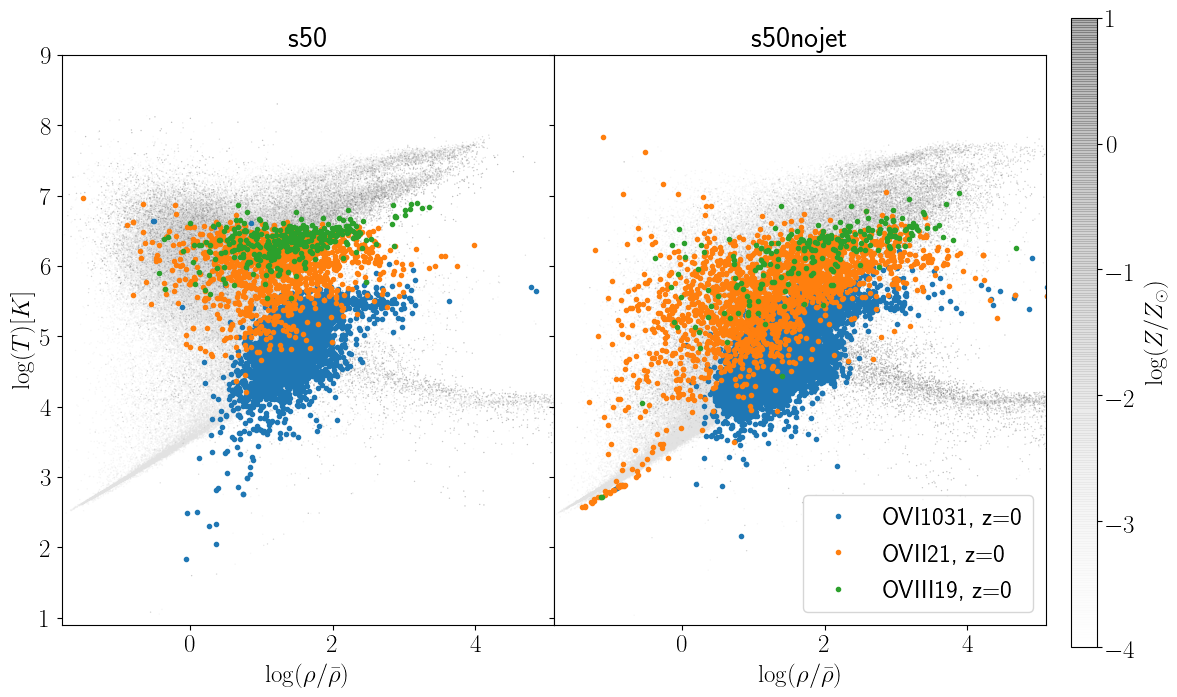}
        \caption{Temperature-baryon overdensity phase space diagrams for the s50 (left) and s50nojet (right) runs. Every 1,000 gas particles (grey-scale) has been plotted, weighted to show the metallicity of the particle normalised by solar metallicity. Each \ion{O}{vi} (blue), \ion{O}{vii} (orange) and \ion{O}{viii} (green) absorber identified has been marked on.}
        \label{fig: Phase Plots}
    \end{figure*}

Figure \ref{fig: Phase Plots} displays overdensity vs. temperature diagrams for the gas in the s50 (left panel) and s50nojet (right) runs at $z=0$. The detected absorption lines are shown for \ion{O}{vi} (blue points), \ion{O}{vii} (orange) and \ion{O}{viii} (green). In the background in grey, a random subset of $0.1\%$ of the gas particles have been plotted, which have been shaded by the metallicity of the particle.

Looking first at the overall gas particles, we see the s50 run has an extension of particles towards high temperature and low density, relative to s50nojet.  As noted in \citet{2020MNRAS.499.2617C}, this can be directly attributed to AGN jet feedback, as a significant fraction of the particles in that region of phase space have been directly ejected and heated by jet mode feedback that entrains IGM gas and pushes it into lower-density regions.  This is the process responsible for the large increase in the WHIM fraction with AGN jets on, from $\sim 30$ to $70\%$~\citep{2020MNRAS.499.2617C, Sorini_2021}, although the bulk of the WHIM is hidden underneath the coloured points.  There is also widespread metal enrichment in both models, even into the low-$T$, low-density spur of particles that represents photo-ionised IGM gas giving rise to the Ly$\alpha$ forest.

Moving on to the absorbers, Figure \ref{fig: Phase Plots} shows that with jets on, the absorbers are detected over more tightly constrained regions of phase space. The three different ions clearly trace different regions in phase space.  This is expected because ionising oxygen up to higher states requires higher temperatures at a given density, and lower densities at a given temperature. As a result, the ions trace roughly distinct regions of temperature space, modulo a mild density dependence.

\ion{O}{vi} (blue points) occurs in both collisionally-ionised gas up to $\sim 300,000$~K where the ion has its collisional ionisation peak, but also photo-ionised gas at $\la 50,000$~K, as suggested by \citet{Oppenheimer_2009}.
\ion{O}{vii} (orange) traces the widest range of cosmic densities, including even in the void regions where Ly$\alpha$ forest gas resides.  \ion{O}{viii} occurs in a narrower and hotter range of temperatures. The large span of densities shows that oxygen is widely distributed in the IGM (even without jets), thus the amount of absorption in a given ion will be set primarily by the amount of gas in the appropriate phase.  Qualitatively, the trends are similar for both s50 and s50nojet, but we will see next that there are some quantitative differences in the physical conditions when jets are turned on.

\subsection{Phase Space Histograms}

    \begin{figure}
        \centering
        \includegraphics[width=\linewidth]{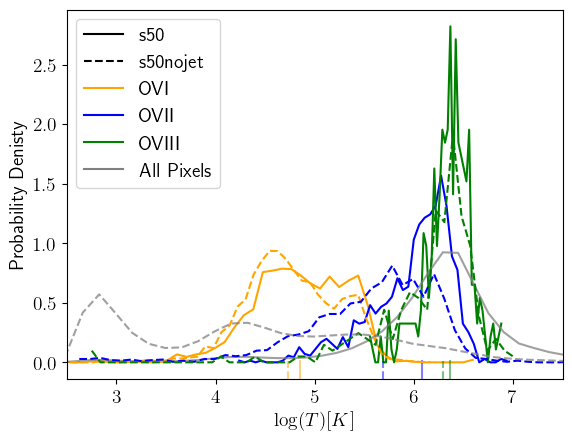}
        \caption{Temperatures of \ion{O}{vi} (orange), \ion{O}{vii} (blue) and \ion{O}{viii} (green) for the s50 (solid lines) and s50nojet (dashed lines) runs. These were generated using 10,000 randomly generated spectra per line at $z=0$. Along the abscissa the median values have been marked on.}
        \label{fig: temp hist}
    \end{figure}

We quantify the trends in the above phase space diagram by constructing histograms in density, temperature, and metallicity for the identified absorbers.  This will illustrate the physical conditions over which each type of ion is prevalent in the IGM.

Figure \ref{fig: temp hist} shows histograms of the temperatures of \ion{O}{vi} (orange), \ion{O}{vii} (blue) and \ion{O}{viii} (green) for the s50 (solid lines) and s50nojet (dashed lines) runs at $z=0$.  Along the abscissa the median values have been marked with vertical ticks.  Note that these histograms have been normalized to be probability densities, and hence they integrate to the same value; a comparison of the raw number of absorbers will be done when we examine the column density distributions in \S\ref{sec: obvs properites}.

The clear separation in temperatures traced by these various ion absorbers is evident.  \ion{O}{vi} traces IGM gas roughly from $T\sim 10^4-10^{5.5}$~K, \ion{O}{vii} from $T\sim 10^{5.3}-10^{6.3}$~K and \ion{O}{viii} traces quite hot gas at $T\ga 10^{6.3-6.7}$~K, albeit with significant overlap.  This indicates that \ion{O}{vii} and \ion{O}{viii} are better direct tracers of the bulk of WHIM gas, leaving aside the issue of observational feasibility~\citep{2017AN....338..281N}.  There are essentially no absorbers found with $T>10^7$~K despite \ion{O}{viii} continuing to have a large ionisation fraction at those temperatures, mostly because gas at those temperatures tends to live in halos which have a fairly small cross-section relative to the bulk of the IGM. Both simulations show gas (as grey points) having $T>10^7$~K at overdensities corresponding to being within bound haloes ($\rho/\bar\rho\ga 100$), but there are very few identified absorbers in that regime among our 10,000 LOS.  This means that it is not typically the case that \ion{O}{vii} and \ion{O}{viii} absorption probes gas within or around massive halos.

Looking at the difference between feedback models (solid and dashed lines), we see that for \ion{O}{vi} the distribution of temperatures remains similar after the the addition of jet feedback, with the median temperature of these detections increasing from $T_{\rm s50nojet}\sim\SI{e4.75}{\kelvin}$ to $T_{\rm s50}\sim\SI{e4.85}{\kelvin}$. Interestingly, both of these median temperatures are not only below the temperatures expected for the WHIM, they are also below the peak ionisation fraction of \ion{O}{vi} in CIE. Hence as mentioned before, \ion{O}{vi} found along random LOS will predominantly arise from photoionisation at a temperature of $T\sim\SI{e4.2}{\kelvin}$. Despite this, Figure \ref{fig: temp hist} still shows non-negligible quantities of \ion{O}{vi} found at CIE temperatures in both runs, with s50 showing a sizeable number of detections at temperatures of $T\sim\SI{e5.5}{\kelvin}$.  It is curious that even this CIE peak is only mildly increased by the large additional amount of WHIM gas in s50 vs. s50nojet.  The reason is that, as pointed out in \citet{2020MNRAS.499.2617C}, the AGN jets take photo-ionised gas and heat it up to $T\ga 10^6$~K, but the fraction of gas around $T\sim 10^{5.5}$~K coincidentally ends up relatively unchanged.  Therefore, in \simba, \ion{O}{vi} seems to be less ideal for exploring the impact of AGN feedback on the WHIM.

The temperature histograms for \ion{O}{vii} show a larger change between different feedback models.  AGN jet feedback introduces the largest change in temperature for \ion{O}{vii} among the various ions.  The histogram with jets on shifts substantially towards higher $T$, driving the median temperatures from $T_{\rm s50nojet}\sim\SI{e5.65}{\kelvin}$ to $T_{\rm s50}\sim\SI{e6.07}{\kelvin}$. Unlike \ion{O}{vi} and \ion{O}{viii} which have relatively narrow CIE temperature peaks, for \ion{O}{vii} the range over which it dominates in CIE is quite broad, which enables it to be sensitive to an overall shift in the WHIM temperatures.  We note that \citet{2018Natur.558..406N} analysed observations of their two \ion{O}{vii} absorbers to obtain estimated absorber temperatures of $6.8^{+9.6}_{3.6}\times 10^5$~K and $5.4^{+9}_{1.7}\times 10^5$~K, which are well within the expected range for both feedback models; \citet{Mathur_2003} likewise estimated $T<10^6$K for their \ion{O}{vii}+\ion{O}{vi} system.  In contrast, \ion{O}{viii} is more like \ion{O}{vi} where the overall histogram is not strongly shifted, and the median temperature is barely affected, going from $T_{\rm s50nojet}\sim\SI{e6.33}{\kelvin}$ and $T_{\rm s50}\sim\SI{e6.37}{\kelvin}$ with the inclusion of jets. 
    
    
    \begin{figure}
        \centering
        \includegraphics[width=\linewidth]{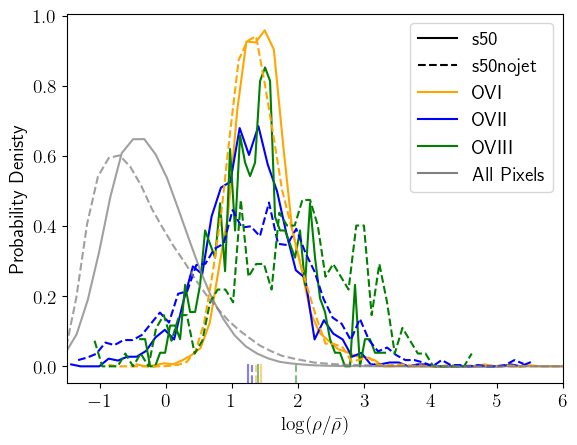}
        \caption{The baryon overdensity of \ion{O}{vi} (orange), \ion{O}{vii} (blue) and \ion{O}{viii} (green). With data attained from 10,000 spectra per line, at $z=0$ for s50 (solid lines) and s50nojet (dashed lines). Along the abscissa the median values are marked on.}
        \label{fig: od hist}
    \end{figure}

We now turn to examining the overdensities of identified absorbers. Figure \ref{fig: od hist} shows histograms of the baryon overdensity for \ion{O}{vi} (orange), \ion{O}{vii} (blue) and \ion{O}{viii} (green) at $z=0$, in the s50 (solid lines) and s50nojet (dashed lines) runs. The median of each data-set has been marked onto the abscissa.
    
Overall, perhaps the most striking result from this is that the overdensity distributions among the three ions strongly overlap, and in most cases have a median around $\rho/\bar\rho \approx 25$, with the one exception being \ion{O}{viii} in s50nojet which shows a median $\rho/\bar\rho \sim 100$.  Thus only for \ion{O}{viii} without jets does the association of these absorbers with gas within or near to groups and clusters hold, but the inclusion of AGN jet feedback in \simba results in much more widespread heating that greatly increases the cross-section of \ion{O}{viii}-absorbing gas in the IGM.  

\ion{O}{vi} and \ion{O}{vii} show minimal changes in overdensities between the feedback models. For both models, \ion{O}{vi} has a sharper distribution, while the tails of \ion{O}{vii} extend to (slightly) higher and lower overdensities.  For \ion{O}{vi} turning on the jets results in the median overdensity increasing from $\rho/\bar\rho \approx 23$ to $\rho/\bar\rho \approx 27$, whereas for \ion{O}{vii} jet feedback the inclusion of jets slightly reduces the median overdensity by $0.06$~dex.  This contrasts with \ion{O}{viii} where jet feedback takes the median value from $\rho/\bar\rho = 112$ to $\rho/\bar\rho = 26$.
It appears that only in the case of \ion{O}{viii} is there a substantial change owing to jets; without widespread jet heating, it appears to be difficult to raise the temperatures of diffuse IGM gas sufficiently to generate much \ion{O}{viii}.  Other that this, all these overdensities broadly correspond to filamentary structures in the Cosmic Web, which if they contained photo-ionised gas would generate \ion{H}{i} column densities of $\sim 10^{14}\cdunits$~\citep{1999ApJ...511..521D}; we will examine the Cosmic Web location of our oxygen lines in \S\ref{sec: web}.

    \begin{figure}
        \centering
        \includegraphics[width=\linewidth]{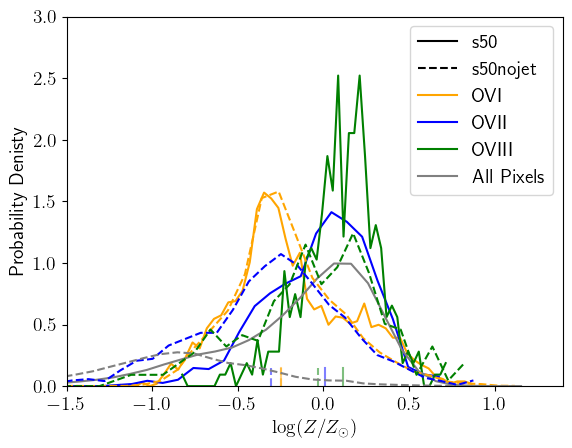}
        \caption{Metallicities of \ion{O}{vi} (orange), \ion{O}{vii} (blue) and \ion{O}{viii} (green) for the s50 (solid lines) and s50nojet (dashed lines) runs. The metallicities are displayed in terms of solar metallicity. These were generated using 10,000 randomly generated spectra per line at $z=0$. Along the abscissa the median values have been marked on.}
        \label{fig: metal hist}
    \end{figure}
    
Finally, we examine the metallicities of absorbing gas.
Figure \ref{fig: metal hist} shows the optical depth weighted metallicities of absorbing gas, relative to solar metallicity~\citep{Asplund_2009}, for \ion{O}{vi} (orange), \ion{O}{vii} (blue) and \ion{O}{viii} (green) for the s50 (solid lines) and s50nojet (dashed lines) runs at $z=0$. Along the abscissa the median values have been marked on.
    
\ion{O}{vi} shows the most consistent behaviour between the two feedback runs, specifically with a detection across the same range of metallicities. The typical values are $Z\sim 0.5Z_\odot$, with a tail to super-solar metallicities.  The \ion{O}{vii} metallicity shows the largest sensitivity to jet feedback, increasing by 0.3~dex up to typically solar metallicity for s50.  This shows that the AGN feedback is transporting significant metals into the WHIM.  \ion{O}{viii} likewise shows an increase, though not as strongly as \ion{O}{vii}, and also lies around solar metallicity.  Recall these metallicity values are weighted by the optical depth in each ion, so they provide the conditions for the gas actually doing the absorption, not typical WHIM gas.

Overall, \simba paints a picture in which high ionisation oxygen absorbers along random LOS arise in a variety of environments, probing from around the cosmic mean density up to warm-hot gas within massive halos. \ion{O}{vi} occurs more often in photo-ionised ($T\la 10^5$~K) versus collisionally ionised gas, and is relatively insensitive to jet feedback (even for collisionally ionised absorbers).  \ion{O}{vii} and \ion{O}{viii} are better direct tracers of the WHIM, spanning a wide range of densities with a median overdensity $\sim 25$ for both absorbers.  The no-jet case is more consistent with the canonical interpretation of \ion{O}{viii} absorbers arising in and around groups and clusters, but with jet feedback, this association is weak and no different than for \ion{O}{vii}.  \ion{O}{vii} can trace temperature shifts owing to jet feedback, but for the other species the constrained ionisation conditions required to produce strong absorption preclude any temperature shift in identified absorbers.

\subsection{Location in the Cosmic Web}\label{sec: web}

    \begin{figure*}
        \centering
        \includegraphics[width=\linewidth]{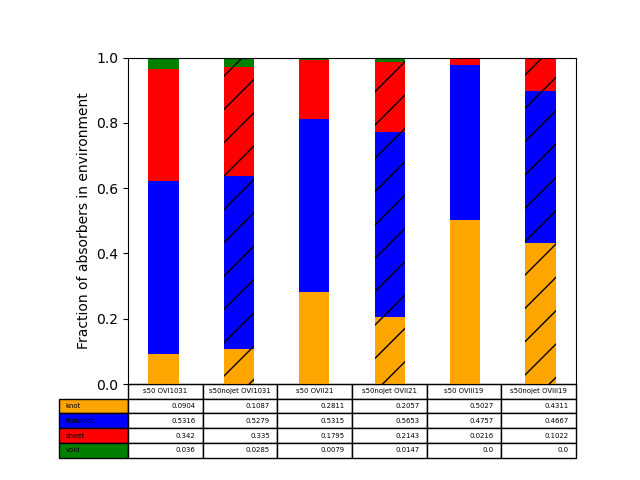}
        \caption{Stacked bar charts of the fraction of knot (orange), filament (blue), sheet (red) and void (green) detections at $z=0$. This was carried out for s50 (plain bars) and s50nojet (diagonally hatched bars) for \ion{O}{vi} (first and second), \ion{O}{vii} (third and fourth) and \ion{O}{viii} (fifth and sixth). The table displays the fractions represented by the bars.}
        \label{fig: Bars}
    \end{figure*}

In the previous section we found that the typical overdensity of high-ionisation oxygen absorbers is relatively modest, broadly corresponding to filamentary structures, but that there is also a large range in overdensities that can give rise to significant absorption.  It is thus worth asking, what are the Cosmic Web environments of these absorbers, in terms of nodes vs. filaments vs. sheets vs. voids?  Also, what sort of halos do these absorbers reside around?  Here, we answer these questions using our simulated absorber sample.

To examine environment, we classify the Cosmic Web location of the absorbers using the \texttt{PWEB} method described in \citet{Cui2018}, which we briefly describe here. Recall that \pygad returns the line of sight and peculiar velocity associated with each absorber, which is converted to a $z$-axis location.  In conjunction with the random $x,y$ coordinates picked for the line of sight, this provides a 3-D absorber position within the simulation volume. 


For each absorber's position, the large scale structure environment in which it was detected was classified with \texttt{PWEB}.
First, the volume of the simulation box is partitioned into a $64^3$ grid, with each cell having a side length of approximately $1.15\textup{Mpc}$. The gravitational potential of each cell is then calculated, and the corresponding Hessian matrix is calculated, given by \citep{2007MNRAS.375..489H}:
    \begin{equation}
        P_{\alpha\beta}=\frac{\partial^2\Phi}{\partial r_{\alpha}\partial r_{\beta}},
    \end{equation}
where $\Phi$ is the gravitational potential, and $P_{\alpha\beta}$ has eigenvalues $\lambda_i$ ($i=1,2,3)$ with $\lambda_1>\lambda_2>\lambda_3$. Finally, the grid cells can be classified as follows:
    \begin{itemize}
        \item void, if $\lambda_1<\lambda_{th}$, 
        \item sheet, if $\lambda_1\geq\lambda_{th}>\lambda_2$,
        \item filament, if $\lambda_2\geq\lambda_{th}>\lambda_3$,
        \item knot, if $\lambda_3\geq\lambda_{th}$,
    \end{itemize}
where $\lambda_{th}$ is a threshold parameter, set to be $\lambda_{th}=0.04$ to roughly match the large-scale matter density distribution. Using this method the $x,y,z$ position of an absorber can be translated into a large scale structure environmental classification.

We note that this classification is being done on a relatively coarse grid, and there can be a range of densities within a single cell.  Thus this analysis is intended to quantify the $\ga$Mpc-scale structures within which these absorbers reside, taking a larger-scale view than the particular overdensities of gas giving rise to absorption that we examined in the previous section.
    

We begin by presenting statistics of which environments host which absorption ions, based on the above classification.  Figure \ref{fig: Bars} shows stack bar charts, and a table, to display the fraction of knot (orange), filament (blue), sheet (red) and void (green) detections at $z=0$. This was carried out for s50 (plain bars) and s50nojet (diagonally hatched bars) for \ion{O}{vi} (first and second), \ion{O}{vii} (third and fourth) and \ion{O}{viii} (fifth and sixth). In the table below the bar graph we list the specific fractions of absorbers arising in each environment.

It seems that all the absorbers are found across all types of structures -- filaments, sheets, and knots.  Generally, the most common environment is filaments (blue shading), but there is also a clear trend that the lowest ion \ion{O}{vi} has a greater contribution from more diffuse environments like sheets, while higher ions preferentially appear more in knots.  This is especially noticeable for \ion{O}{viii}, which is actually more common in knot (or node) regions than filaments.  These trends are as expected, as the denser regions should contain hotter gas that results in higher ionisation.  The weakness of the environmental trends is somewhat surprising, however, since it is often believed that \ion{O}{vi} arises almost exclusively in filaments and sheets while \ion{O}{viii} arises mostly in knots.  While broadly true, our analysis suggests that such associations are not so clearly delineated.

We can also investigate the impact that AGN jets have on the environments of absorbers, by comparing adjacent solid (s50) and hatched (s50nojet) bar graphs.  \ion{O}{vi} shows minimal change in absorber environments, with a very small shift away from knots when turning on the jets.
In contrast, for \ion{O}{vii} the difference in fractions between s50 and s50nojet is significantly larger, and the inclusion of jets is showing a clear movement of absorbers towards the knots, at the expense of the other environments.  
\ion{O}{viii} shows a similar increase in knot absorbers, at the expense of sheet absorbers.  Thus the overall trend is that AGN jets tend to increase the high-ionisation absorption in the knot regions. This might be expected because the jets themselves are coming from the most massive galaxies which tend to reside in such knots.

    
    \begin{figure*}
        \centering
        \includegraphics[width=\linewidth]{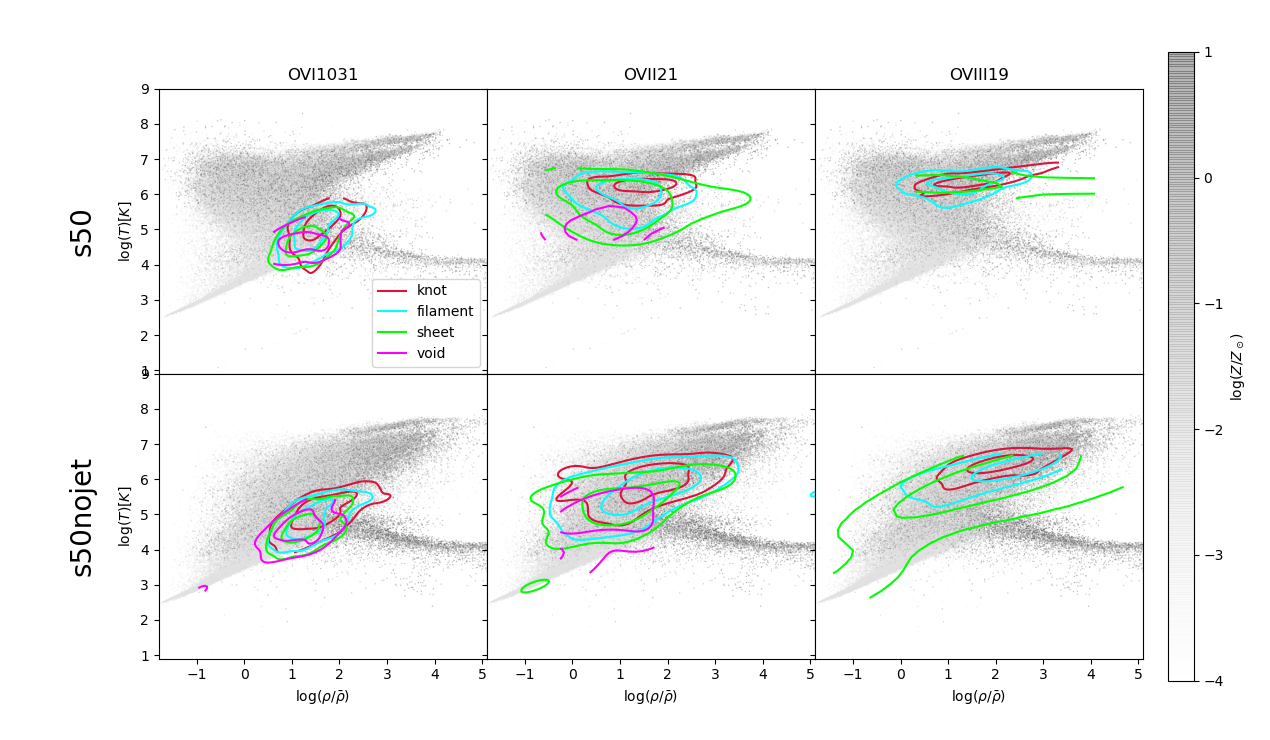}
        \caption{Temperature-baryon overdensity phase space diagrams for the s50 (top row) and s50nojet (bottom row) runs. Every 1,000 gas particles (grey-scale) has been plotted, weighted to show the metallicity of the particle normalised by solar metallicity. For \ion{O}{vi} (left), \ion{O}{vii} (middle) and \ion{O}{viii} (right) contours containing $50\%$ and $90\%$ of the detected absorbers in each environment have been added. Knot (red), filament (cyan), and sheet (green) detections occurred for all three ions, but there were no void (magenta) detections for \ion{O}{viii}.}
        \label{fig: Contours}
    \end{figure*}

To explore absorber environments in more detail, we can view the location of each type of environment within a cosmic phase space diagram. Figure \ref{fig: Contours} shows the cosmic phase diagrams of a random sample of 0.1\% of \simba's gas particles (grey-scaled by metallicity), marked with contours containing $50\%$ and $90\%$ of the \ion{O}{vi} (left), \ion{O}{vii} (middle) and \ion{O}{viii} (right) absorbers detected within the knot (red), filament (cyan), sheet (green), and void (magenta) \texttt{PWEB} environments. These absorbers are from our 10,000 mock spectra at $z=0$ in the s50 (top row) and s50nojet (bottom row) runs. Note that there are no void \ion{O}{viii} absorbers in either simulation, so the magenta lines don't appear in the rightmost column.

There is substantial overlap in the phase space location of absorbers between the various environments.  This occurs because the large-scale structures within which absorbers live can be quite different, even when the local physical conditions required to give rise to strong absorption in a particular ion are similar.  Nonetheless, some general trends are noticeable, in that void absorbers tend to be skewed towards slightly lower overdensities, while sheet absorbers span a very wide range in overdensities.  Knot absorbers tend to have the highest temperatures and typically somewhat higher overdensities than filament absorbers.  Hence some trends reflecting expectations that knots should lie at the densest locations in the web while the voids should occupy the least dense are reflected in these trends, albeit with large overlap.

Comparing the full physics s50 run (top row) versus s50nojet (bottom row), we see that \ion{O}{vi} shows the least variations among environments of the three ions.  Nonetheless, the s50nojet clearly shows a larger range of overdensities probed within each environment, and thus overall.  There are several 
s50nojet void detections in the $-1\lesssim\log(\rho/\bar{\rho})\lesssim0$ regime, where none appeared in s50.  The temperature range spanned is not significantly separated by environment; both photo- and collisionally ionised absorbers occur in all environments.
    
For \ion{O}{vii}, knot detection in s50 are constrained to higher temperatures than in s50nojet, and all occur within the WHIM range.  Therefore dense large-scale environments with \ion{O}{vii} mostly probe $T>10^6$~K with jets, where in s50nojet they can probe to lower temperatures. 
One of the most noticeable differences between the s50 and s50nojet results arises in the contours of the sheet detections. In the s50nojet run we found a significant number of \ion{O}{vii} absorbers around $-1\lesssim\log(\rho/\bar{\rho})\lesssim-0.5$, $T\sim\SI{e3}{\kelvin}$, i.e. gas typically associated with the Ly$\alpha$ forest, that were not present in the s50 run. This can be seen more clearly in Figure \ref{fig: Phase Plots}. In both runs the void contours became less constrained, this was due to the small number of absorbers from this environment detected. Less than $2\%$ of the \ion{O}{vii} absorbers detected were from this environment. Despite this, we still see that the s50nojet run is still being detected at lower minimum temperatures. 
    

\subsection{Distance to Nearest Galaxy Halo}
    
    \begin{figure}
        \centering
        \includegraphics[width=\linewidth]{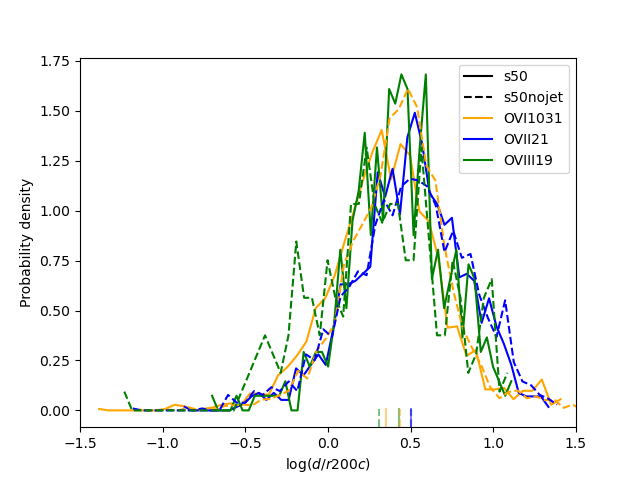}
        \caption{Histogram of the distance to the nearest central galaxy, normalised by $r200c$, of \ion{O}{vi} (orange), \ion{O}{vii} (blue) and \ion{O}{viii} (green) absorbers. The absorbers were detected at $z=0$ in the s50 (solid) and s50nojet (dashed) runs. Along the abscissa the median values have been marked on.}
        \label{fig: distance hist}
    \end{figure}

A complementary approach to characterising environments is to examine the nearest galaxy, in projection.  This approach is often used by observers to associate absorption with a particular galaxy that the metals potentially originated from. Unfortunately, such a correspondence can be problematic, as the metals giving rise to high-ionisation oxygen absorbers are typically injected a long time ago~\citep[``ancient outflows", in the lingo of][]{Ford_2014}.  Nonetheless, measuring the distance to the nearest galaxy can be an interesting approach to characterising absorber environments.

Figure \ref{fig: distance hist} shows histograms of the minimum projected normalised distance to a central galaxy at $z=0$ for the s50 (solid lines) and s50nojet (dashed lines), where the normalisation is by its halo's virial radius defined as enclosing 200 times the critical density ($r_{200c}$). Along the abscissa the median values for \ion{O}{vi} (orange), \ion{O}{vii} (blue) and \ion{O}{viii} (green) have been marked on.

All absorbers predominantly arise well outside the virial radii of galactic halos, typically at $\sim 2-3 r_{200c}$.  Strikingly, there is not a large difference in the distance to nearest galaxies among the different ions; the histograms to first order all overlap.  An exception is the case of \ion{O}{viii} for s50nojet, which we have seen before arises in higher overdensity gas, and correspondingly shows a smaller distance to nearest galaxy.  Nonetheless, corroborating the cosmic web classification above, none of these ions in either model arise predominantly within galactic halos.

    
    
    
The broad conclusion from our environments study is that all these ions tend to arise in diffuse regions of the cosmic web outside of halos, predominantly within filamentary large-scale structures.  The differences between the typical environments of these various absorbers is modest, with only a mild trend that higher ionisation lines tend to arise in denser regions of the cosmic web.  This trend occurs without jets, but is exacerbated slightly by the action of jets which tends to heat the gas closer to the denser parts of large-scale structure.
AGN jet feedback had a larger impact on the physical conditions of \ion{O}{vii} and \ion{O}{viii} absorbers, with jet feedback increasing the probability of high temperatures and metallicities, as well as generated some low density detections.  

\subsection{Global Evolution of Oxygen Ions}\label{sec: evolution}

Among the motivating factors for studying the WHIM are the inferences that can be made about galaxy evolution from the metal content of the WHIM. To further explore this, we traced the evolution of the global mass density $\Omega$ of our WHIM tracer ions in each feedback model from $z=3$ to $z=0$.  We scaled this mass density to the present-day critical density as follows:
    \begin{equation}
        \Omega_{ion}(z) = \frac{\rho_{cm}(z)}{h^3(1+z)^3\rho_{crit,0}}
    \end{equation}
where $\rho_{cm}$ is the comoving mass density, $h=0.68$ is the Hubble constant and $\rho_{crit,0}=(3H_0^2/8\pi G)$ (\citealt{2020ARA&A..58..363P}). This was done for \ion{O}{vi}, \ion{O}{vii}, and \ion{O}{viii} as well as all oxygen in the s50 and s50nojet runs.  We computed individual particle ionisation fractions using \texttt{PYGAD}, which provides this functionality that is also used in the spectral generation.  We summed over the particle's ion masses and divided by the simulation box's comoving volume and required constants to give $\Omega_{ion}$. For the total oxygen the same process was followed, with no ionisation correction applied.  Note that \simba directly tracks elemental oxygen production (and locking into stars and dust) based on supernova yields.
    

\begin{figure} 
    \centering
    \includegraphics[width=\linewidth]{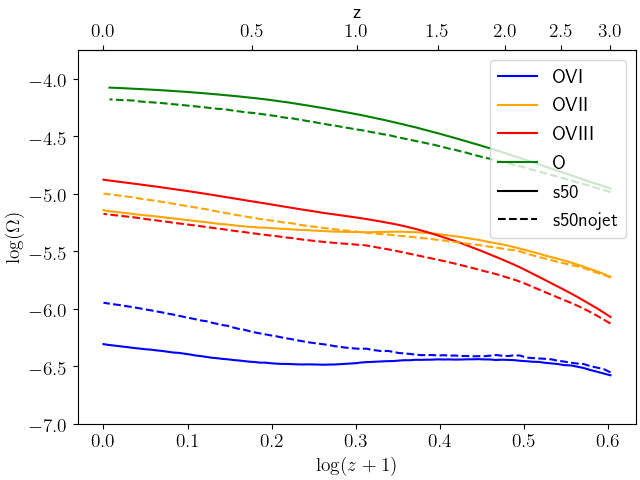}
    \caption{Evolution of the cosmic mass density $\Omega$ of total oxygen (green), \ion{O}{vi} (blue), \ion{O}{vii} (orange) and \ion{O}{viii} (red), in s50 (solid line) and s50nojet (dashed line), as a function of redshift from $z=0-3$.}
    \label{fig: Omega O}
\end{figure}
        
Figure \ref{fig: Omega O} shows the evolution of these ion densities, as a function of $\log(z+1)$, with the corresponding values of $z$ shown for convenience along the top axis. The evolution of $\Omega$ for \ion{O}{vi} (blue), \ion{O}{vii} (orange), \ion{O}{viii} (red), and total oxygen (green) is tracked across a redshift range of $z=3$ to $z=0$ for s50 (solid lines) and s50nojet(dashed). 

The total O abundance grows with time, and the ion densities generally do so as well, but not always.  \ion{O}{vi} shows a nearly flat evolution in the s50 model, only rising slightly from $z\sim 0.5$ to $z=0$.  \ion{O}{vii} shows a similar behaviour, though there is more of a noticeable rise at early redshifts.  \ion{O}{vii} most faithfully tracks the overall O abundance.

Comparing s50 vs. s50nojet, the overall oxygen abundance is lower in s50nojet.  Although there are more stars and hence metals produced in s50nojet (since this run does not quench massive galaxies), the amount of oxygen locked into stars is also significantly greater, thus leaving slightly less oxygen within the gas phase globally.  \ion{O}{viii} again mimics the trend in total oxygen, with an exaggerated difference between the models owing to the impact of jet feedback heating placing more gas at \ion{O}{viii}-absorbing temperatures.  Meanwhile, \ion{O}{vi} and \ion{O}{vii} show little difference in their $\Omega$ values until $z\la 1$. At earlier epochs, the black holes in \simba are not typically in jet mode \citep{2019MNRAS.487.5764T}, because this is only activated at low Eddington rations that tend to occur in massive galaxies with little cold gas appearing at later cosmic epochs.  
    
At low redshift the jet feedback had the most sizeable impact on the \ion{O}{vi} mass density, with \ion{O}{vi} is suppressed by $\sim 0.4$~dex at $z\la 0.5$ owing to the jet feedback.  Jets more modestly suppress \ion{O}{vii} by $\sim 0.2$ dex at late epochs.  These oxygen atoms are at least partly in the \ion{O}{viii} phase which increases by $\sim 0.3$ dex, though it is already enhanced at this level by $z\sim 1$. Overall, this shows that the impact of jet feedback on absorption mostly begins at $z\la 1$ and is strongest at $z\la 0.5$.

The behaviour exhibited by \ion{O}{vi} and \ion{O}{viii} is primarily due to the heating effects of the jets. As previously mentioned, the jets are capable of reaching velocities of $\SI{8000}{\kilo\metre\per\second}$. Even with retardation from gas interactions and gravity over cosmic time periods the heating effects of the jets will reach out to mega-parsec scales (\citealt{2020MNRAS.491.6102B}). Furthermore, in the runs with no jet feedback, gravitational shocks heat regions around filamentary structures, however this heating does not have the extent of the jet feedback. Hence, for many regions in s50nojet with temperatures $T\sim\SI{e5}{\kelvin}$, comparable regions in jet runs will have temperatures of $T\sim10^6-\SI{e7}{\kelvin}$ (\citealt{2020MNRAS.499.2617C}). As previously discussed, the ionisation fraction of \ion{O}{vi} peaks at $T\sim\SI{e5.5}{\kelvin}$, decreasing rapidly on either side of the peak. Consequently, with the jet feedback reducing the volume of regions around this peak temperature this explains the suppression of \ion{O}{vi} in jet runs. Conversely, the peak fraction of \ion{O}{viii} peaks at the upper limits of the WHIM temperature range, and hence the increased volume of regions with these temperatures gives rise to the higher \ion{O}{viii} abundance at low redshift.  
    
The relatively consistent behaviour seen for \ion{O}{vii} is expected owing to the helium-like structure of this ion, creating a plateau in its ionisation fraction with respect to temperature. Thus for a region at $T\sim\SI{e5.5}{\kelvin}$ in a no jet run, a comparable region in a jet run with $T\sim\SI{e6.5}{\kelvin}$, would show minimal change in the abundance of \ion{O}{vii}. 
    
Overall, as the simulation evolves the impact of AGN feedback increases. Not only is the jet feedback increasing the fraction of baryons in the WHIM \citep{2020MNRAS.499.2617C} at $z\la 1$, it is pushing them to higher temperatures. Subsequently, the jet feedback reduces the quantity of the baryons that could be traced using \ion{O}{vi}, but increases the quantity that could be traced by \ion{O}{viii}, with \ion{O}{vii} being the least affected of the high oxygen ions. We now examine how these physical properties of high ionisation oxygen in the IGM can be traced using absorption line statistics.

\section{Column Density Distributions of WHIM Tracers}\label{sec: obvs properites}

In this section, we explore the column density distributions of the WHIM tracers \ion{O}{vi}, \ion{O}{vii}, and \ion{O}{viii}. This provides a test of the predictions from the \simba simulation in comparison with the latest observations, as well as exploring the sensitivity of these predictions to the inclusion of jet feedback, along with the redshift evolution that could be explored with larger samples from future UV and X-ray telescopes.

\ion{O}{vi} is a relatively well-studied ion at low redshifts, with {\it Hubble's} Cosmic Origins Spectrograph Guaranteed Time Observing program~\citep{Danforth_2016} amassing a substantial sample at $0.1< z<0.73$ that provides a stringent test of \simba and its input physics.  For \ion{O}{vii}, the data is much more sparse, with a small handful of confirmed intergalactic absorbers from \citet{2018Natur.558..406N} to which we will make preliminary comparisons. Finally, we make predictions for \ion{O}{viii} for which no definitively intergalactic absorbers are yet known but are hopefully forthcoming with next-generation X-ray facilities.


We focus on the column density distribution as the basic counting statistic for characterising IGM absorbers.  The column density distribution functions (CDDF) is the bivariate distribution of absorbers as a function of column density and redshift. The CDDF is defined as 
\begin{equation}\label{eq: f(N,z)}
    f(N,z) = \frac{\partial^2n}{\partial N\partial X},
\end{equation}
where $N$ is the column density, $n$ is the number of absorbers and $dX$ is the comoving path length defined as
\begin{equation}\label{eq: dX}
    dX = dz\frac{H_0}{H(z)}(1+z)^2,
\end{equation}
where $z$ is redshift, $H(z)$ is the Hubble constant at $z$ and $H_0$ is the Hubble constant at $z=0$. At a redshift of zero, $dX=dz$ \citep{1969ApJ...156L...7B}, but $dX$ accounts for changes in the CDDF arising due to the cosmological evolution of the path length \citep[see e.g.][]{2019MNRAS.488.2947W}, so that a static comoving absorber population will have the same $f(N,z)$ at all redshifts.


To identify absorbers, we follow the method in \citet{2018MNRAS.477..450N}.  First, we use {\tt yt} and its extension Trident\footnote{https://trident.readthedocs.io/en/latest/index.html} \citep{2017ApJ...847...59H} to make a $1000\times 1000$ pixel map of the projected column density over the whole area of the simulation box, as shown in Figure~\ref{fig: maps}. We use projected maps rather than out Voigt profile fitting results because, particularly for \ion{O}{vi} and \ion{O}{viii}, the absorbers are very sparse, so we would need to generate and fit an enormous sample of LOS to probe the column density distribution over the range that covers the observations.  The column density of the given ion was integrated along the $z$ axis, and projected onto a grid. In this scenario, $dz$ from equation \ref{eq: dX} is the redshift-space depth of the simulation box (e.g. $\Delta z=0.0333564$ at $z=0$) times $1000^2$ LOS.  The CDDF is then just the histogram of these column densities, divided by the bin size in $\log N$.  The CDDFs generated in this way from \simba are shown in Figure~\ref{fig: OVI comp}, which will be discussed in the following subsections.

\subsection{\ion{O}{vi} CDDF}

\begin{figure*} 
    \centering
    \includegraphics[width=0.9\linewidth]{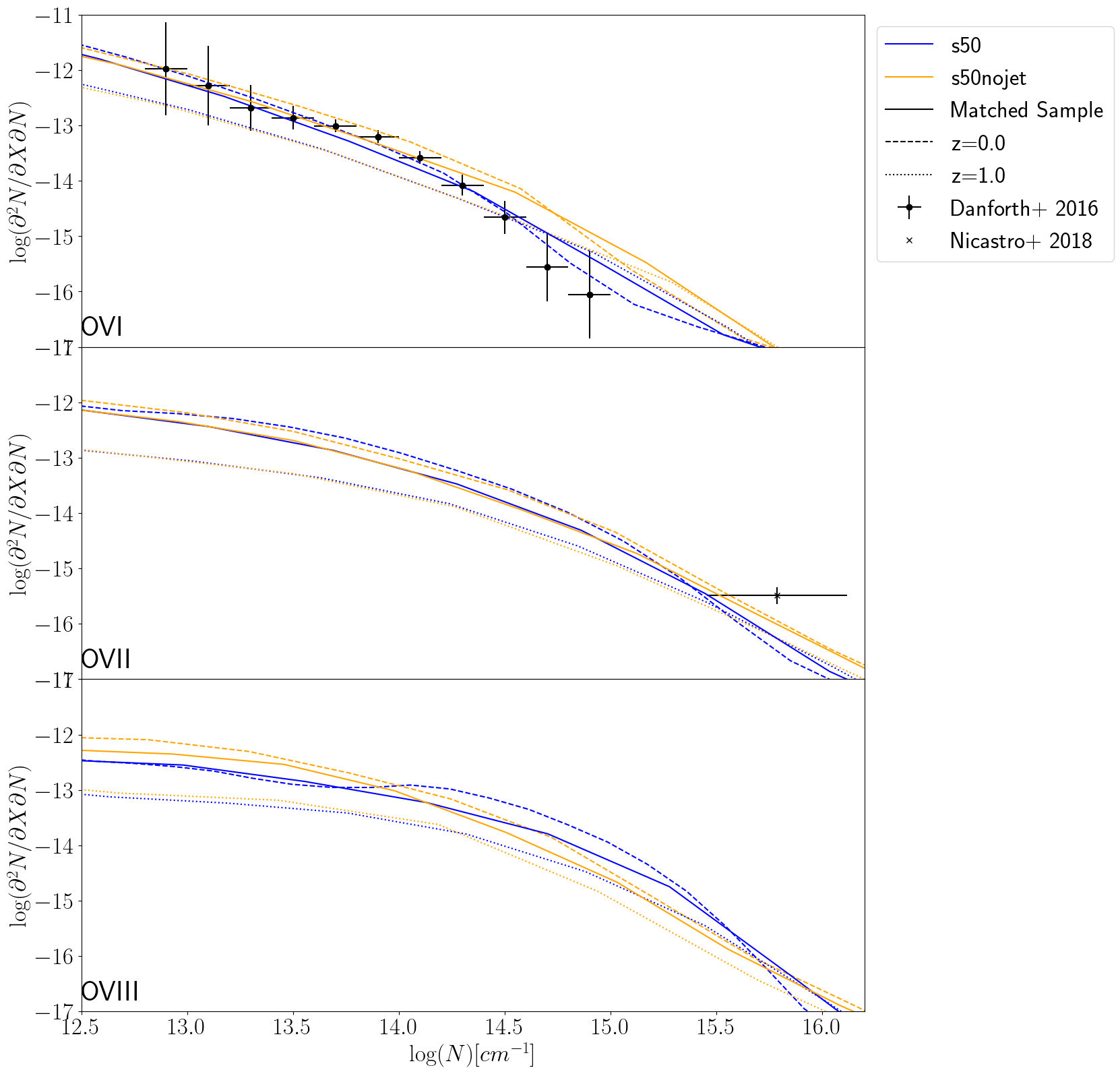}
    \vskip-0.1in
    \caption{CDDF a in the s50 (blue) and s50nojet (orange) runs for \ion{O}{vi} (top panel) \ion{O}{vii} (middle) and \ion{O}{viii} (bottom).  The dashed and dotted lines show $z=0$ and $z=1$, respectively. The solid line shows the results from samples matched in redshift distribution to observations (black points) from  \citet{Danforth_2016} for \ion{O}{vi} ($\bar{z}\approx 0.28$) and \citet{2018Natur.558..406N} for \ion{O}{vii} ($\bar{z}\approx 0.4$), as described in the text; there is no data available for \ion{O}{viii} so its solid line is shown at the same redshift as \ion{O}{vii}.  All CDDFs increase from $z=1\to 0$, particularly at the low-$N$ end.  s50 and s50nojet are essentially identical at $z=1$, but show increasing differences to lower redshifts. The agreement with observations is generally good (comparing to the solid lines), though in detail the \ion{O}{vi} appears mildly over-predicted at the high-$N_{\rm OVI}$ end and under-predicted for \ion{O}{vii}. }
    \label{fig: OVI comp}
\end{figure*}

Figure \ref{fig: OVI comp}, top panel, shows the CDDF from \simba's s50 (blue) and s50nojet (orange), compared to the HST-COS data obtained from \citet{Danforth_2016} (black points).  We obtained the observations from the HST MAST archive for all the COS-GTO data, and constructed a sample of lines identified as \ion{O}{vi}(1032\AA) by the COS-GTO team.  The redshift path length of $\Delta z=14.49$ was taken from \citet[Table 4]{Danforth_2016}.  In order to account for redshift evolution when comparing to observations, we created a mock sample that matches the redshift distribution of observed absorbers. We generate column density maps of the twenty snapshots corresponding to the redshift range $0.1\la z\la 0.73$, select a fraction of LOS from among all $10^6$ LOS that matches the fraction of the \ion{O}{vi} absorbers in each of those redshift bins, and combine those to make our predicted CDDF.  We note that the median redshift of the COS-GTO sample of \ion{O}{vi} is 0.28, and the CDDF constructed at this redshift from the projected map method is not significantly different from the redshift-matched mock sample.  Voigt profile fitting yields a mildly lower CDDF, but still generally within uncertainties.

Comparing the solid line to the \citet{Danforth_2016} data, there is good agreement between the simulated and observed CDDFs across the full column density range, particularly for the full \simba run (s50).  In detail, s50nojet tends to slightly over-predict the high-$N_{\rm OVI}$ end.  This is a non-trivial success of the \simba model, showing that the widespread heating of the IGM by \simba's jet feedback if anything improves agreement with observations of high-ionisation IGM metal absorbers~\citep[as also seen for the \lya\ forest;][]{2020MNRAS.499.2617C}.  However, it is also the case that s50 seems to under-predict the CDDF at intermediate column densities, where the observations show a ``bump" in the CDDF at $N_{\rm OVI}\sim 10^{13.8}\cdunits$ that is not reproduced in either \simba run (or any other simulation that we are aware of).

The redshift evolution from $z=1$ to $z=0$  shifts the predicted \ion{O}{vi} CDDF upwards with time, particularly at the lower column densities.  At $z=1$, there is no difference between the s50 and s50nojet CDDFs, which indicates that the heating of IGM gas to the temperatures required for \ion{O}{vi} is as yet dominated by gravitational processes.  Going to lower redshifts, AGN jet feedback adds substantial heat to the IGM, which causes oxygen to be pushed into higher ionisation states and thus dampens the CDDF evolution rate.  As such, by $z=0$ the \ion{O}{vi} has increased substantially more than for the s50 run.  The difference is particularly noticeable at the high-$N_{\rm OVI}$ end, which is arising from the filamentary structures that are most subject to AGN jet heating.  

Compared to other recent simulations, \simba provides comparable or better agreement with observations.  Over most of the range, \simba yields similar predictions to IllustrisTNG~\citep{2018MNRAS.477..450N}, but at the highest observed columns  IllustrisTNG  tends to be closer to the no-jet results. The error bars on the observations are large here, so these discrepancies should not be over-interpreted, nonetheless it is interesting that better constraints there could potentially provide discrimination between models.  \citet{2018MNRAS.477..450N} also found that the original Illustris simulation substantially under-predicted the \ion{O}{vi} CDDF at most $N_{\rm OVI}$, showing that agreement with these data is non-trivial.  Likewise, the EAGLE simulation significantly under-predicted the COS-GTO \ion{O}{vi} data, though it agreed better with previous smaller samples taken using the Space Telescope Imaging Spectrograph instrument~\citep[see][for details]{2016MNRAS.459..310R}. These trends are consistent with the amount of heating each simulation's feedback model produces outside of galaxies: \simba the most, IllustrisTNG somewhat less, and EAGLE the least.  This mirrors the trend  in the amount of evacuation of hot gas from galaxy groups seen amongst these simulations~\citep{Oppenheimer_2021}. While all these simulations match well at low-$N_{\rm OVI}$, it appears that significant IGM heating and enrichment such as that provided from AGN feedback in \simba or IllustrisTNG seems to result in better agreement with \ion{O}{vi} observations at the high-column end.

\subsection{\ion{O}{vii} CDDF}

Figure \ref{fig: OVI comp}, middle panel shows the CDDF for \ion{O}{vii} at $z=0$ (dashed), $z=0.406$ (solid), and $z=1$ (dotted) for the s50 (blue) and s50nojet (orange) runs.  We choose the intermediate as the closest snapshot to the mean redshift of the two detected IGM \ion{O}{vii} absorbers in \citet{2018Natur.558..406N}.  A CDDF constructed from these two absorbers is shown as the data point with errors.  We use a $dz=0.42$ as quoted in \citet{2018Natur.558..406N}, and assume a bin of $10^{15-16}\cdunits$.  With only two absorbers, the bin size is somewhat arbitrary, and was chosen to comfortably encompass the two detections with a lower limit set by the approximate detection threshold.  The minimum span to encompass the absorbers would be $\Delta N_{OVII}=4.4-7.8\times 10^{15}\cdunits$, which would increase the inferred CDDF by 0.4~dex, so one should regard the vertical (statistical) errorbar shown on the data point as sub-dominant to a larger systematic error that is difficult to quantify and hence not shown.

There is strikingly little differences between the predictions of two AGN feedback models at any redshift.  This is somewhat surprising, since the IGM physical conditions have changed significantly due to jets with more than double the WHIM gas, and the morphology of the \ion{O}{vii} absorption looks significantly different in the projected maps in Figure~\ref{fig: maps}. This likely arises because while some \ion{O}{vi} absorbing gas is pushed into the \ion{O}{vii} regime by jets, a similar amount is pushed from \ion{O}{vii} into \ion{O}{viii}.  Thus, it appears that coincidentally, the \ion{O}{vii} CDDF is fairly insensitive to \simba's jet feedback.

Differences between the models do appear at the highest columns ($N_{\rm OVII}\ga 10^{15}\cdunits$).  This is where the  \citet{2018Natur.558..406N} data lies, and we see that both runs tend to have difficulty producing enough high-$N_{\rm OVII}$ absorbers to reach the observed CDDF.  This was also the case for EAGLE as shown in \citet{2018Natur.558..406N}, although other models (not fully hydrodynamic galaxy formation simulations) could match their data.  Indeed, even if one combined the \ion{O}{vi} and \ion{O}{viii} CDDFs and assumed those lines were actually all \ion{O}{vii} lines, it would still not reach the value of the observations.  This suggests a more fundamental failing in \simba, such as not enough oxygen being produced overall. However, \simba's good agreement with the galaxy mass-metallicity relation, as well as the \ion{O}{vi} CDDF, suggests that \simba's overall metal production is reasonable.

In detail, the s50nojet run is closer to the observations, so it appears that jet heating of the IGM tends to push more \ion{O}{vii} into \ion{O}{viii} and hence worsen agreement with observations.  It is worth noting that our modelling assumes ionisation equilibrium, but particularly in very diffuse gas non-equilibrium effects can become important, which could potentially increase  \ion{O}{vii} absorption~\citep{Oppenheimer_2013}.
With only two data points, one should be cautious about reading too much into these disagreements, but overall it seems that high-$N_{\rm OVII}$ absorbers that will hopefully be detected with upcoming facilities such as {\it Athena} and {\it Lynx} can help constrain AGN feedback models and their impact on the IGM.

Meanwhile, it could be that the observations are biased in some way, or that it is simply small number statistics with only two absorbers.  It is possible that the estimate of $N_{\rm OVII}$ in \citet{2018Natur.558..406N} is too high, as this was done via photoionisation modeling that required some assumptions. However, we have also checked directly versus \simba's predicted equivalent width distribution, and the discrepancy is comparable.  Finally, it could be that the particular blazar (1ES 1553+113), chosen purely because it is X-ray bright, just happens to probe a region well suited for \ion{O}{vii} absorption.  We note the authors find a bright galaxy just 129~kpc away from one of the absorbers, putatively within its halo's virial radius, which is closer in than typical \ion{O}{vii} absorbers we find in \simba.  Thus it may be that one these strong absorbers is the result of a fortuitous choice of sight line.  However, the other absorber does not have any comparably nearby bright galaxy.  Thus overall, the discrepancy is not easy to explain, and it will be interesting to see if forthcoming X-ray facilities confirm the high level of absorption inferred by this XMM data. 

\subsection{\ion{O}{viii} CDDF}

Figure \ref{fig: OVI comp}, bottom panel, similarly shows the \ion{O}{viii} CDDF.  The line types correspond to the redshifts as in the middle panel. There is no observational comparison available here, so only the predictions for the s50 (blue) and s50nojet (orange) are shown. 

\ion{O}{viii} shows differences even at $z=1$, unlike the lower ionisation lines.  Already by this epoch, the AGN jet feedback has caused significant heating that reflects in more \ion{O}{viii} absorption at $N_{\rm OVIII}\ga 10^{14.5}\cdunits$.  This difference grows stronger at lower redshifts.  This shows that the impact of AGN heating is most readily observable in \ion{O}{viii}, among these high-ionisation oxygen lines.  This is likely because it is more difficult to achieve the temperatures required for \ion{O}{viii} by gravitational shock heating on filaments, so the effects of AGN feedback are more prominent. 

At $z=0$, the s50 CDDF shows a curious bump at $N_{\rm OVIII}\sim 10^{14-14.5}\cdunits$, which is not seen in s50nojet.  This appears to be real, as we have sliced the data in many ways and seen a similar feature each time.  It is possible that this reflects the column density range down to which the IGM has been heated via jet feedback.  Temperature slices from \simba as shown in \citet{2020MNRAS.499.2617C} indicate a clear boundary out to which heating extends, and if such heating is contributing significantly to the \ion{O}{viii} abundance, then this is an interesting prediction.  However, we caution that this \simba run's $50\hmpc$ volume may not be large enough to fully capture the full extent of IGM heating owing to a lack of high-mass halos, so the feature may be at least partly an artefact of numerical parameters.

\citet{Bonamente_2016} reported the detection of two putative intergalactic \ion{O}{viii} absorber at $z=0.09$, roughly associated with a broad \lya\ line and with some \ion{O}{vi} lines in the vicinity.  The column densities are estimated to be $5\times 10^{17}\cdunits$ and $6\times 10^{16}\cdunits$, with large systematic uncertainties.  Given the small path length of this $z=0.177$ quasar, and \simba's predicted rapid drop of the \ion{O}{viii} CDDF at high-$N_{\rm OVIII}$, it is clear that the simulations cannot come close to the implied CDDF value of this data.  However, it is difficult to interpret this observation statistically, because the LOS was chosen based on having \ion{O}{vi} and broad \lya\ lines, so likely probes a biased region of the Universe.  Their preliminary estimates indicate that the absorber has a temperature of $T\ga 10^{6.4}$~K and occurs at an overdensity $\ga 100$, which are also higher than typical values for \ion{O}{viii} from \simba, again suggesting that the detected absorbers are not representative.  Hence we refrain from drawing any conclusions from comparisons to these observations.

Overall, while there are notable discrepancies between \simba predictions and observations of the \ion{O}{vi}, \ion{O}{vii}, and \ion{O}{viii} CDDFs, \simba generally does a reasonable job of reproducing the available data, comparable to or better than other current cosmological galaxy formation simulations.  The greatest constraining power for AGN feedback models appears to be provided by the upper end of the column density distributions ($N\sim 10^{15-16}\cdunits$), which is 
encouraging because this will be the region of parameter space that is most straightforward to probe with upcoming X-ray telescopes.  It is also possible that a comprehensive archival search for \ion{O}{vi} among all available {\it Hubble} ultraviolet quasar spectra could populate the high-$N$ end of the \ion{O}{vi} CDDF, although this may be a complicated task owing to heterogeneous selection and varying sensitivities.  It is beyond the scope of this work to engage in such a project, but our findings offer hope that high-ionisation IGM oxygen absorbers could provide interesting constraints on AGN feedback in the near future.

\section{Conclusions}\label{sec: conculusion}

We have examined intergalactic absorption in three key high-ionisation metal ions, \ion{O}{vi}, \ion{O}{vii}, and \ion{O}{viii}, within the \simba cosmological hydrodynamic simulation~\citep{2019MNRAS.486.2827D}. \simba includes unique models for black hole growth and feedback, in particular its bipolar jet-mode AGN feedback has been demonstrated to have widespread effects in distributing and heating intergalactic baryons~\citep{2020MNRAS.491.6102B,2020MNRAS.499.2617C, Sorini_2021}.  In this work we explore the physical conditions of these high-ionisation absorbers, quantify their environments, and compare their statistics to available observations.  To illustrate the sensitivity to AGN feedback, we compare the full-physics \simba model (s50) versus a model run with identical initial conditions but not including jet or X-ray AGN feedback (s50nojet); both are run within $50\hmpc$ volumes, not \simba's fiducial $100\hmpc$ volume since this does not have a no-jet analogue.  Appendix A shows that the X-ray mode has negligible impact on the IGM absorbers considered in this work, and the differences are driven by \simba's jet AGN feedback.  Our main conclusions are as follows:

\begin{itemize}
    \item The heating from jet feedback in \simba is clearly noticeable along individual lines of sight and in the projected column density maps, becoming more dramatic for higher ionisation ions.  This suggests that high-ionisation metal absorbers have the potential to constrain AGN feedback models in a completely new regime from the galaxy and black hole properties to which they are usually tuned.
    
    \item Using a newly-developed Voigt profile fitter to associated absorbing gas with the underlying line-of-sight physical conditions at $z=0$, we find that \ion{O}{viii} probes the hottest IGM gas at a typical $T\sim 10^{6.3}$~K, \ion{O}{vii} probes $T\sim 10^{6}$~K gas, and \ion{O}{vi} probes $10^5$~K gas with a wider spread.  This is as expected based on the collisional ionisation fractions peaks of these ions.
    
    \item In contrast, there is no strong trend of the gas density probed versus ionisation state, with all ions typically probing moderate overdensities of $\delta\sim 20-30$; the exception is \ion{O}{viii} in the no-jet run which probes $\delta\sim 100$.  The ions also probe a similar range of overdensities, from $\sim 1-1000$.  This suggests that one should be cautious in uniquely associating high-ionisation lines, even \ion{O}{viii}, with warm-hot gas in halos or their outskirts; instead, it appears that these absorbers are able to more generally trace the filamentary Cosmic Web.
    
    \item The optical depth-weighted metallicity of the absorbing gas ranges from typically $\sim 1/2$ solar for \ion{O}{vi} to roughly solar for the higher ions. The optical-depth weighting means that it is likely to be biased high relative to the typical gas metallicity found at these filamentary overdensities.  This distinction is important when using photoionisation models to infer the physical properties of such absorbers in observations.
    
    \item Using PWEB to characterise the Cosmic Web environment, we find that all absorbers are present in all environments except voids, though they are most likely to be found in filaments (consistent with the typical overdensities).  There is a modest but clear trend that higher ionisation lines correspond more closely to knots, while \ion{O}{vi} tend to be stronger in sheets.  The trends are subtle, again precluding a strong association of Cosmic Web environment with oxygen ionisation state.
    
    \item A similar story is found when examining the $r_{200c}$-scaled distance to the nearest halo: There is no clear trend with ionisation state, and most absorbers are found outside of halos, typically at $2-3r_{200c}$ but with a wide range. Combined with the large-scale environment information, \simba predicts that most absorers are located outside of the halos in the filament environment. 
    
    \item The redshift evolution of the global IGM mass density in each ion indicates that all oxygen ions broadly increase in comoving mass density with time tracking the overall oxygen abundance.  However there are clear trends such as \ion{O}{vi} showing rather flat evolution from $z=3$ to $z=0$, while \ion{O}{viii} increases the most rapidly.  This reflects the fact that heating the IGM to \ion{O}{viii} requires either large shocks or AGN feedback, both of which are more common towards lower redshifts.  In general, s50 produces more IGM oxygen density in metals than s50nojet likely owing to it ejecting more metal-enriched gas from halos, which is reflected in \ion{O}{vi} and \ion{O}{viii}, but is slightly inverted for \ion{O}{vii} at low redshifts.  These trends highlight the complex interplay between enrichment, transport, and ionisation level of IGM gas in setting the absorption properties.
    
    \item We examine how these redshift evolution and feedback differences manifest in the observable column density distribution functions (CDDFs) for these ions, computed from projected ion density maps.  At $z=1$, little difference is seen between s50 and s50nojet, except for a slightly higher \ion{O}{viii} CDDF for the s50 model.  This reflects the relatively minimal amount of AGN jet heating in the IGM by $z\ga 1$ in \simba's model.
    
    \item By $z=0$, the CDDFs of the full \simba model for \ion{O}{vi} is clearly lower for s50, and higher for \ion{O}{viii}.  This reflects a general shift towards higher temperature absorption owing to AGN heating.  Interestingly, the \ion{O}{vii} shows little difference at any redshift between the two models, suggesting that while IGM heating ionises some \ion{O}{vi} to \ion{O}{vii}, a similar amount is ionised up from \ion{O}{vii} to \ion{O}{viii}.
    
    \item Comparing to \ion{O}{vi} observations from the COS-GTO team~\citep{Danforth_2016}, we find that a redshift-matched sample of \simba absorbers provides a good match to the observed \ion{O}{vi} CDDF.  This agreement is non-trivial, as some simulations have failed to match this data (generally falling too low) though others like IllustrisTNG match as well as \simba.  There are small discrepancies at intermediate columns in s50 and at high columns in s50nojet that may offer paths forward to constrain AGN feedback models with improved statistics.
    
    \item There are few confirmed intergalactic \ion{O}{vii} absorbers due to the line falling in the soft X-ray band, but comparing to the two strong absorbers seen by \citet{2018Natur.558..406N} suggests that \simba has difficulty producing enough \ion{O}{vii} absorbers, with s50nojet being only slightly closer than s50.  Since the two absorbers arise along a single LOS, cosmic variance may as yet be large, so we refrain from drawing firm conclusions about AGN feedback from this.  Nonetheless, the differences in the predictions at $N_{\rm OVII}\ga 10^{15.5}\cdunits$ offers hope that the increased statistics provided by future X-ray observatories may prove to be a powerful constraint on models.
    
    \item The \ion{O}{viii} CDDF shows a strong a strong increase with redshift, more so for the no-jet model.  The s50 model at $z=0$ shows a curious bump that may arise from the division between jet-heated and non-jet-heated IGM.  While the differences between the s50 and s50nojet runs are modest, they are most apparent at $N_{\rm OVIII}\ga 10^{15}\cdunits$ that may be accessible with upcoming X-ray facilities.
    
\end{itemize}

One of the goals of this work was to test whether the strong widespread heating by AGN jet feedback in \simba could be ruled out with existing high-ionisation IGM metal absorbers.  It appears that not only can it not be ruled out, in some aspects it provides a better agreement with available data than a model without jets.  This suggests that \simba's prediction of a dominant fraction of the present-day IGM being in the WHIM remains viable in comparison with current observations. Such observations seem to be on the cusp of providing significant constraints on AGN feedback models, motivating future facilities in both the UV and X-ray, particularly in the X-ray where the discovery space remains quite open. In the near future, fast radio bursts represent another promising technique for probing the low-redshift IGM \citep[see, e.g.,][]{Lee_2021}, and thus may offer additional constraints for feedback models implemented in simulations such as \simba. In forthcoming work, we aim to make more tailored forecasts and predictions for upcoming observatories, and further explore the physical processes connecting black holes on sub-pc scales deep within galaxies with the Mpc-scale diffuse intergalactic medium.

\section*{Acknowledgements}

We acknowledge helpful discussions with Sarah Appleby, Jacob Christiansen, and Dylan Robson.  We thank Philip Hopkins for making \gizmo\ public, Horst Foidl, Thorsten Naab and Bernhard Roettgers for developing and maintaining \pygad, and J. Xavier Prochaska, Nicolas Tejos and Joe Burchett for {\sc PyIGM}.  RD acknowledges support from the Wolfson Research Merit Award program of the U.K. Royal Society.
Throughout this work, DS was supported by the European Research Council, under grant no. 670193, and by the STFC consolidated grant no. RA5496. WC is supported by the STFC AGP Grant ST/V000594/1. He further acknowledges the science research grants from the China Manned Space Project with NO. CMS-CSST-2021-A01 and CMS-CSST-2021-B01.

\simba was run on the DiRAC@Durham facility managed by the Institute for Computational Cosmology on behalf of the STFC DiRAC HPC Facility. The equipment was funded by BEIS (Department for Business, Energy \& Industrial Strategy) capital funding via STFC capital grants ST/P002293/1, ST/R002371/1 and ST/S002502/1, Durham University and STFC operations grant ST/R000832/1. DiRAC is part of the National e-Infrastructure. 

\section*{Data and Software Availability}

The simulation data underlying this article are publicly available at \url{https://simba.roe.ac.uk}. The software used in this work is freely available at the URLs posted in the text, and the derived data will be shared on request to the corresponding author.

\bibliographystyle{mnras}
\bibliography{high_ionisation_Oxygen_in_SIMBA}

\appendix
\section{Feedback comparisons}
        
\begin{figure} 
    \centering \includegraphics[width=\linewidth]{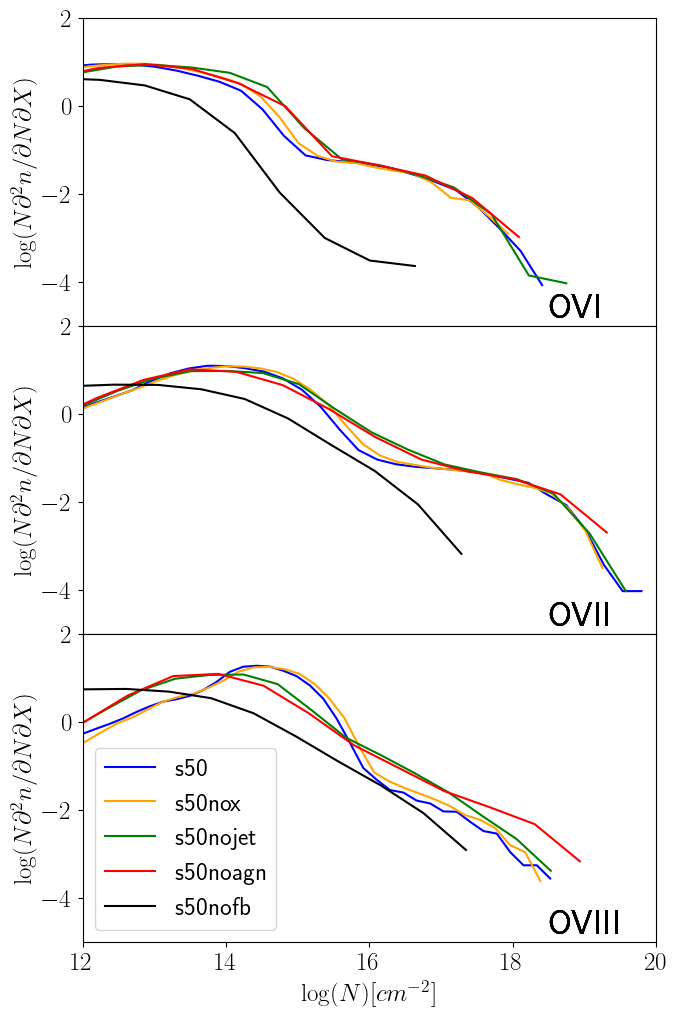}
    \caption{The column density distribution functions multiplied by the column density ($Nf(N,z)$) of \ion{O}{vi} (left), \ion{O}{vii} (middle) and \ion{O}{viii} (right), for all four AGN feedback models, s50 (blue), s50nox (orange), s50nojet (green), s50noagn (red) at $z=0$. The differences between models are not dramatic, but in general the s50 and s50nox runs are quite similar, and the s50nojet and s50noagn models are also similar.  This indicates that X-ray and radiative AGN feedback are of lesser importance relative to jet AGN feedback in setting high-ionisation IGM absorption properties.}
\label{fig: CDDF}
\end{figure}

As mentioned earlier, the \simba suite contains feedback variants with different feedback modules turned off.  The s50 run contains the full \simba physics, s50nox turns off X-ray AGN feedback, s50nojet additionally turns off jet-mode AGN feedback, and s50noagn additionally turns off radiative-mode AGN feedback.  Finally, s50nofb further turns off star formation driven winds, and thus it has no kinetic feedback of any sort.

Figure \ref{fig: CDDF} shows 
$Nf(N,z)$ as a function of column density $N$, where $f(N,z)$ is the CDDF. We multiply by $N$ in order to better visually differentiate the CDDFs of the various runs.  The panels show \ion{O}{vi} (left), \ion{O}{vii} (middle) and \ion{O}{viii} (right) at a redshift of $z=0$. The four different AGN feedback models are displayed for each ion s50 (blue), s50nox (orange), s50nojet (green) and s50noagn (red), as described in \S\ref{sec: SIMBA}.  These are shown over a wider range in column densities than displayed in Figure~\ref{fig: OVI comp}, to explore the differences more fully.
        
The differences for \ion{O}{vi} and \ion{O}{vii} are rather modest, showing that there is broadly limited sensitivity to AGN feedback in these ions, as also discussed in \S\ref{sec: obvs properites}.  The largest differences occur for \ion{O}{viii}.  Here, it is clearly seen that s50 and s50nox are quite similar, while s50nojet and s50noagn are distinct from these and similar to each other.  Going from s50noagn to s50nojet differs purely from including radiative AGN feedback, while going from s50nox to s50 differs purely from including X-ray AGN feedback.  The effects of these modes are clearly not as important as AGN jet feedback, which is isolated by examining s50nojet versus s50nox.  This motivates our study in the main text focusing primarily on s50 vs. s50nojet.

There are some other interesting differences.  At very high column densities $N_{\rm OVIII}\ga 10^{17}\cdunits$, it appears radiative AGN feedback is having a large effect.  These absorbers are extremely rare, but offer a way to probe feedback associated with the main growth mode of AGN.  More subtly, we can see that s50 tends to have the lowest \ion{O}{vi} CDDF among the models at low columns (where differences are noticeable).  Also, there is the interesting cusp in each CDDF in most the models, at differing column densities for each ion; this is only clearly noticeable if one plots $Nf(N,z)$ rather than $f(N,z)$.  It is not entirely clear what causes this feature, but it does appear that AGN feedback generally exacerbates it, particularly for the higher ions.

\bsp
\label{lastpage}
\end{document}